\documentclass[11pt,a4paper]{article}
\pdfoutput=1
\usepackage{jheppub}

\def\nn{\nonumber}

\def\bea{\begin{eqnarray}}
\def\eea{\end{eqnarray}}
\def\dq{{[dQCD]}}
\def\sc{{[SCET]}}

\def\as{\alpha_s}

\def\figscale#1#2{\pdfximage width#2 {#1.pdf}\pdfrefximage\pdflastximage}

\title{Quantifying Comparisons of Threshold Resummations}
\author{George Sterman and Mao Zeng}
\affiliation{C.N.\ Yang Institute for Theoretical Physics and Department of Physics and Astronomy\\
 Stony
Brook University, Stony Brook, New York 11794--3840, USA}

\emailAdd{george.sterman@stonybrook.edu}
\emailAdd{mao.zeng@stonybrook.edu}

\abstract{We explore similarities and differences between widely-used threshold resummation formalisms, employing electroweak boson production as an instructive example.  Resummations based on both  full QCD and soft-collinear effective theory (SCET) share common underlying factorizations and resulting evolution equations.    The formalisms differ primarily in their choices of boundary conditions for evolution, in moment space for many treatments based on full QCD, and in momentum space for treatments based on soft-collinear effective theory.  At the level of factorized hadronic cross sections, these choices lead to quite different expressions.   Nevertheless, we can identify a natural expansion for parton luminosity functions, in which SCET and full QCD resummations agree for the first term, and for which subsequent terms provide differences that are small in most cases.    We also clarify the roles of the non-leading resummation constants in the two formalisms, and observe a relationship of the QCD resummation function $D(\as)$ to the web expansion.}

\arxivnumber{1312.5397}

\begin{document}
\maketitle
\flushbottom

\section{Introduction}
\label{sec:Intro}

Threshold resummations organize singular distributions in the short distance functions of factorized inclusive hadronic cross sections to all orders in perturbation theory.  This organization has been carried out in a variety of ways.   Early work on the subject identified the importance of soft gluon radiation under phase space restrictions, and derived leading logarithmic corrections, including the role of the running coupling \cite{Parisi:1979xd,Curci:1979am,Chiappetta:1982mw}.   Threshold resummations including all logarithmic orders were based on the factorization of partonic cross sections into field-theoretic matrix elements in Ref.\ \cite{Sterman:1986aj} and on direct diagrammatic analysis  in Refs.\ \cite{Catani:1989ne, Catani:1990rp}.    Since then, many developments and applications employing these approaches have appeared, including those described in Refs.\ \cite{Contopanagos:1996nh, Catani:1998tm, Korchemsky:1992xv,Korchemsky:1993uz,Belitsky:1998tc,Catani:2001ic,Catani:2003zt}, which we will have occasion to use below.   These resummations were carried out directly in perturbative QCD, an approach that is sometimes referred to as ``full QCD", and which we will call direct QCD (dQCD) below.

Much recent discussion has concerned the apparent contrast between dQCD threshold resummations \cite{Sterman:1986aj, Catani:1989ne, Catani:1990rp, Contopanagos:1996nh, Catani:1998tm, Catani:2003zt} and those based on 
effective field theory  \cite{Becher:2007ty, Ahrens:2008nc,Bauer:2010vu,Bauer:2010jv}, in particular on soft-collinear effective theory (SCET)   \cite{Bauer:2000yr,Bauer:2001yt,Bauer:2000ew}.     In important analyses, Refs.\ \cite{Bonvini:2012az,Bonvini:2013td} and \cite{Becher:2007ty,Ahrens:2008nc,Becher:2006mr} confirmed the analytic equivalence of SCET and dQCD methods at various levels, and highlighted differences based on differing scale choices and treatments of non-singular behavior at threshold.    Our aim here is to extend these analyses and to provide a further perspective on the relationship between  direct QCD and SCET resummations, by pointing out their common basis in the factorization properties of soft gluon radiation \cite{Contopanagos:1996nh}.    This enables us to derive their key formulas side-by-side, and in this way to clarify their similarities and differences.   We will also show that in practical terms, both approaches should give nearly identical predictions for many cross sections, even though at intermediate stages their treatments of hard-scattering functions differ markedly.  For example, once the resummed cross sections are each computed in a manner that avoids extraneous differences in nonsingular terms, their numerical predictions are essentially indistinguishable for Higgs boson production at LHC kinematics.     These conclusions follow from a new expansion of the cross section based on the shape of parton luminosities, which may have further applications.

Both dQCD and SCET threshold resummations use Mellin or Laplace transforms as a tool, employing much the same all-orders analysis developed for the resummation of logarithms in transverse momentum using impact parameter space \cite{Collins:1981uk}.   In direct QCD, the factorized cross sections are often given as  integrals of products of Mellin- or Laplace-transformed hard scattering functions and parton distribution functions.   In SCET, transforms may appear only as aids in solving evolution equations, and resummed cross sections are given directly in momentum space as convolutions of hard scattering functions and parton distributions.   As we shall see, however,  both hard-scattering functions are derived from the same evolution equation, applied to the same ``soft function", the vacuum expectation value of a product of Wilson lines \cite{Korchemsky:1992xv,Korchemsky:1993uz,Belitsky:1998tc}.

We emphasize at the outset that the specific direct QCD and SCET formalisms that we compare are by no means unique.   For definiteness, we follow the presentations of \cite{Catani:2003zt} and \cite{Becher:2007ty} for dQCD and SCET respectively, but the conclusions we reach should apply to other  applications of threshold resummation in QCD and SCET.    Further clarifying the common and distinct features of threshold resummation in the notation of these well-known papers will lead us to a number of new results as well.\footnote{Many of the same considerations apply to resummed event shape distributions, although based on factorizations that are different from the threshold case \cite{Almeida:2014uva}.}

Our discussion in Sec.\ \ref{sec:soft} begins  with a review to set notation for the cross sections under consideration.  We go on to recall the re-factorization of soft radiation near partonic threshold \cite{Sterman:1986aj, Becher:2007ty, Korchemsky:1994is}, and the consequent evolution equations, common to both direct QCD and SCET.   We recall the differing choices of boundary conditions in the direct QCD and effective theory formalisms, and the corresponding expressions for the resummed partonic and hadronic cross sections.   No new results are derived in this section, but we believe the parallel development of the central results of the two formalisms, with an independent treatment of factorization and other scale dependences, clarifies their relationship.   

Section \ref{sec:webevolve} reformulates evolution in terms of the logarithm of the renormalized soft function.  This logarithm, or ``resummed exponent" has a direct interpretation in terms of diagrams called webs \cite{Sterman:1981jc,Gatheral:1983cz,Frenkel:1984pz}.  We compare treatments in moment and momentum space, applying renormalization group equations to the resummed exponent \cite{Collins:1981uk,Contopanagos:1996nh}. We focus in particular on the non-leading resummation function commonly denoted $D(\as)$, which along with the cusp anomalous dimension determines all logarithms at partonic threshold.    The function $D(\as)$  turns out to be proportional to the momentum-space sum of the web diagrams, evaluated at a specific scale.  

In Sec.\ \ref{sec:partonic} we extend the results of Ref.\ \cite{Bonvini:2013td}, and show how the effective theory and direct QCD agree for partonic cross sections, up to subleading logarithmic differences in moment space, when the soft scale  is chosen for this purpose.   We will see that this agreement follows systematically at any order from the evolution equation for the sum of web diagrams, derived in Sec.\ \ref{sec:webevolve}.  Finally, in Sec.\ \ref{sec:hadronic} we find that, despite their different treatments of evolution for the soft function, it is actually natural to anticipate agreement for direct QCD and soft collinear effective theory resummations of hadronic cross sections, for most of the interesting parameter range involved in collider experiments.  We identify a natural expansion of the parton luminosity function in its logarithmic derivatives, whose first term gives dQCD and SCET resummations that agree identically for a natural choice of the SCET soft scale.   We include two appendices that deal with technical developments.

\section{Factorization, re-factorization and evolution}
\label{sec:soft}

We begin by setting the notation for Drell-Yan-like, color-singlet inclusive cross sections, including the cases of electroweak vector boson and Higgs production.    We go on to recall the factorization of soft radiation near partonic threshold, identify the resulting evolution equation, and contrast the solutions of this equation in dQCD moment-space resummation \cite{Sterman:1986aj, Catani:1989ne, Catani:1990rp, Contopanagos:1996nh, Catani:1998tm, Catani:2001ic,Catani:2003zt} and the soft-collinear effective theory (SCET) momentum space resummation method as developed in Refs.\ \cite{Becher:2007ty, Ahrens:2008nc}.  

\subsection{Electroweak annihilation and its moments}
\label{sec:ewa}

 For the production of a system of a large mass $M$ by electroweak annihilation, the cross section is given in factorized form by 
\begin{eqnarray}
   \frac{d\sigma_{AB\rightarrow M}(S,M^2)}{dM^2} &=& \sum_{{\rm partons}\ a,b} 
    \int dx_a\,dx_b\
    f_{a/A}(x_a,\mu_f)\,f_{b/B}(x_b,\mu_f)
    \nn\\
	&\   & \hspace{20mm}  \times \hat\sigma_{ab \rightarrow M} \left(M^2,\hat s, M^2/\mu_f^2,\alpha_s(\mu_f)\right)
    \nonumber\\	
    &\ & \hspace{-20mm} =
	     \sum_{{\rm partons}\ a,b} \     \sigma_0^{a b}(S,M^2)\ 
	 \int \frac{dx_a}{x_a}\,\frac{dx_b}{x_b}\
     f_{a/A}(x_a,\mu_f)\,f_{b/B}(x_b,\mu_f)\;  
     \nn\\
     &\ & \hspace{20mm} \times\
    C_{ab \rightarrow M} \left(z, M^2/\mu^2_f,\alpha_s(\mu_f)\right)
    	\nonumber\\	&\ & \hspace{-20mm} =
\sum_{{\rm partons}\ a,b} \     \sigma_0^{a b}(S,M^2)\ 
	 \int_\tau^1 \frac{dz}{z}\ {\cal L}_{ab}\left( \frac{\tau}{z},\mu_f\right)\ C_{ab \rightarrow M} \left(z, M^2/\mu^2_f,\alpha_s(\mu_f)\right), \nonumber \\
\label{eq:basic-fact}
\end{eqnarray}
where $\hat \sigma_{ab\rightarrow M}$ is a perturbative quantity that begins with the lowest order (LO) cross section, $ \sigma_0^{ab}(\hat{s},M^2)$,
  in QCD, with $\hat s \equiv x_a x_b S$, 
and where $ C_{ab\rightarrow M+X}$ is a dimensionless ``hard-scattering" or ``coefficient function", which begins at order unity for the annihilation processes allowed by the parton model. For the Drell-Yan process, $\{ab\} = \{q,\bar q\}$ or $\{\bar q,q\}$, while for Higgs production via gluon fusion, $\{ab\}=\{g,g\}$.   (With a much smaller contribution from light quarks, $\{a,b\}=\{q\bar q\}$.)  
For such inclusive cross sections, the hard-scattering functions $C_{ab}$ are also in convolution with differential partonic luminosities, ${\cal L}_{ab}$ defined by
\begin{eqnarray}
{\cal L}_{ab}\left( \frac{\tau}{z},\mu_f\right) = z\ \int \frac{dx_a}{x_a}\,\frac{dx_b}{x_b}\
     f_{a/A}(x_a,\mu_f)\,f_{b/B}(x_b,\mu_f)  \, \delta\left( z - \frac{\tau}{x_ax_b}\right)\, .
\label{eq:Ldef}
\end{eqnarray}
For simplicity, we have suppressed the hadronic labels $A,B$ here and below in the luminosity.
 In the above expressions, we recall a conventional notation,
\begin{align}
\tau &\equiv \frac{M^2}{S}\, ,
\nn\\
z &\equiv \frac{\tau}{x_ax_b}\, ,
\label{eq:ztau-def}
\end{align}
in terms of which we can identify true threshold, $S \rightarrow M^2$, and partonic threshold, $x_ax_bS \rightarrow M^2$, at $\tau=1$ and $z=1$, respectively.   

The hard-scattering functions, $C_{ab}$ in Eq.\ (\ref{eq:basic-fact}), are generally singular at partonic threshold, and
threshold resummation organizes singular distributions for $z\rightarrow 1$.   
These distributions are organized by moments with respect to $\tau$ at fixed $M$, under which the cross sections (\ref{eq:basic-fact}) factorize into products, 
\begin{align}
\int_0^1 d\tau \, \tau^{N-1}  \frac{1}{ \sigma_0^{a\bar a}(S,M^2)}\ \frac{d\sigma^{(a)}_{AB\rightarrow M}(S,M^2)}{dM^2}
& \nn\\
&\ \hspace{-45mm} =\ \sum_{a \leftrightarrow \bar a} 
 \widetilde f_{a/A}(N,\mu_f)\, \widetilde f_{\bar a/B}(N,\mu_f)\
 \widetilde C_{a \bar a \rightarrow M} \left(N, M/\mu_f,\alpha_s(\mu_f)\right)
 \nn\\
&\ \hspace{-45mm} =\ \sum_{ a \leftrightarrow \bar a} 
 \widetilde {\cal L}_{a\bar a}(N,\mu_f)\
 \widetilde C_{a\bar a \rightarrow M} \left(N, M/\mu_f,\alpha_s(\mu_f)\right)\, .
 \label{eq:moment-product}
 \end{align}
The moments of parton distributions with respect to the $x_i$ are given by
 \begin{align}
  \widetilde f_{c/C}(N,\mu_f) = \int_0^1 dx\, x^{N-1} f_{c/C}(x,\mu) \, ,
  \label{eq:Mellin-def}
 \end{align}
 and analogously for the functions $\widetilde {\cal L}$ and $\widetilde C$.  As indicated by the notation in Eq.\ (\ref{eq:moment-product}), and the following, with Higgs and Z production in mind, we will simplify by assuming that only a single parton-antiparton combination is relevant, which we will generally denote by $a,\bar a$.   The cross section $d\sigma^{(a)}/dM^2$ then represents the partonic combination that requires threshold resummation, and the notation $a\leftrightarrow \bar a$ in Eq.\ (\ref{eq:moment-product}) indicates the exchange of the roles of quark and antiquark, when applicable.   
  
  The inverse Mellin transform
  from  $\widetilde g(N)$ to $g(x)$, with $g$ any
  of these functions, is given by
  \begin{align}
  g(x) = \int_{{\cal C}_g} \frac{dN}{2\pi i}\, x^{-N}\ \widetilde g(N)\, ,
  \label{eq:mellin-inversion}
  \end{align}
  where the contour ${\cal C}_g$ in the complex $N$ plane is to the
  right of all the singularities of $g(N)$.   This implies that $g(x)=0$ for $x>1$.   
  Under the Mellin transform, or a closely-related Laplace transform introduced below,
  logarithmic singularities of the form $\ln^m(1-z) / (1-z)$ transform into series of
  logarithms in $N$ beginning with $\ln^{m+1}N$, and similarly for the inverse transform.

\subsection{Soft gluon re-factorization and the soft function}
\label{sec:re-fact-soft}

The essential property of the hard-scattering functions $C_{a \bar a}(z)$, 
which makes resummation possible, is a re-factorization of
singular behavior for $1-z \rightarrow  0$, \cite{Sterman:1986aj,Korchemsky:1992xv,Korchemsky:1994is,Becher:2007ty,Bauer:2010vu} 
\begin{align}
 C_{a\bar a \rightarrow M} \left(z, M^2/\mu^2,\alpha_s(\mu)\right)\ &=  H_{a\bar a} \left(M/\mu,\as(\mu) \right) 
 \nn\\
& \hspace{5mm} \times\ S_{a\bar a}\left (1-z,\frac{M(1-z)}{\mu},\as(\mu)\right) 
 + {\cal O}\left((1-z)^0\right)\, ,
 \label{eq:refactorization}
\end{align}
where as indicated, corrections are less singular than $1/(1-z)$ or $\delta(1-z)$ as $z\rightarrow 1$.   Such corrections, which may include powers of $\ln(1-z)$, contribute at  order $1/N$ times logs of $N$
\cite{Laenen:2008ux}.  The factorization scale, labelled $\mu$ here can be chosen equal to the renormalization scale for this discussion.
The quantity $S_{a\bar a}$ is a ``soft function", whose field theoretic definition, given below, is essentially equivalent in full QCD and soft-collinear effective theory.    

Once a general factorization theorem, Eq.\ (\ref{eq:basic-fact}) is established, 
the re-factorization, Eq.\ (\ref{eq:refactorization}) of the hard-scattering functions follows in dQCD from the use of the Ward identities of the theory \cite{Sterman:1986aj}, or equivalently in effective field theory language, by a field redefinition for hard collinear quanta \cite{Becher:2007ty,Bauer:2010vu,Bauer:2001yt}.   
Singular $1/(1-z)$ behavior is only present for short-distance functions that describe annihilation, $b=\bar a$, because
all other combinations require the emission of a soft quark or antiquark into the final state, which suppresses infrared behavior.  
We also notice in passing that we are using a conventional, if slightly non-intuitive, terminology in which the hard-scattering function is the product of the soft function with a  short-distance function.   The soft function, however, is determined by perturbative methods, as we now describe.

The soft function, $S_{a \bar a}$, which is termed the ``eikonal hard-scattering function" in Ref.\ \cite{Laenen:2000ij},  is
a dimensionless quantity constructed from the vacuum expectation of a Wilson loop that describes the
annihilation of two lightlike Wilson (or eikonal) lines,\footnote{We have omitted factors of $\sqrt{z}$ found in \cite{Becher:2007ty} because we concentrate on leading $(1-z)$ behavior.}
\begin{align}
S_{a \bar a}\left (1-z,\frac{M(1-z)}{\mu},\as(\mu)\right) 
&= M\,  W_{a \bar a}\left ({M(1-z)},{\mu},\as(\mu)\right) \, .   
\label{eq:soft}
\end{align}
The function  $W_{a\bar a}(\omega,\mu,\as(\mu))$ in turn is defined as  the Fourier transform 
\begin{align}
\label{eq:WDYdef}
  W_{a \bar a}\left (\omega,{\mu},\as(\mu)\right)
   &= \int_{-\infty}^\infty\frac{dx^0}{4\pi}\,
   e^{i\omega x^0/2}\,\widetilde W_{a \bar a} (x^0,\vec{x}=0,\mu) \,,
   \end{align}
   of a Wilson loop vacuum expectation \cite{Sterman:1986aj,Korchemsky:1992xv,Korchemsky:1994is,Becher:2007ty}
   \begin{align}
\label{eq:softmatrixelement}
   {\tilde W}_{a \bar a}(x,\mu) &= \frac{1}{N_c}\,
   \langle 0|\,\mbox{Tr}\,{\bf \bar T}
   \big[ \Phi_{n}^{(a)}{}^\dagger(x) \Phi^{(a)}_{\bar n}(x) \big]\,
   {\bf T} \big[ \Phi^{(a)}_{\bar n}{}^\dagger(0) \Phi^{(a)}_n(0) \big]|0 \rangle \, .
\end{align}
In these matrix elements, the operators $\Phi_\beta$ are lightlike path-ordered exponentials in the directions $\beta$,
\begin{equation}
\label{eq:Phi-n}
   \Phi_\beta^{(a)}(x) = {\rm\bf P}\,\exp\left( -  ig\int_{-\infty}^0\!ds\,\beta \cdot A^{(a)}(x+s\beta) \right)\, .
\end{equation}
These ordered exponentials are matrices in the representations appropriate to the partons $a\bar a$ that annihilate, with $\beta=n,\bar n$ moving in opposite directions, $n \cdot \bar n =1$.

The matrix elements $W_{a\bar a}(x_0\mu)$ can be taken as the starting point for threshold resummation.
In the effective field theory treatment, $W_{a\bar a}$ is evaluated using the SCET ultra-soft gluon field $A_s$ rather than the usual QCD gluon field $A$, but since the Feynman rules involved are the same, the two definitions are  essentially identical.

For all treatments of threshold resummation, Mellin, Eq.\ (\ref{eq:Mellin-def}), and/or Laplace transformations are useful, and equivalent to leading power in $N$ or $1/(1-z)$.  The Laplace transform of the soft function is
\begin{align}
\widetilde S \left( \ln \frac{M}{\bar N\mu}, \as(\mu) \right) &= 
\int_0^{\infty} \frac{d\omega}{M} \, \exp \left[-\frac{N\omega}{M} \right] S\left (\frac{\omega}{M},\frac{\omega}{\mu},\as(\mu)\right)
\nn \\
&= \int_0^1 dz \, z^{N-1} S \left (1-z,\frac{M(1-z)}{\mu},\as(\mu)\right) \ +\ {\cal O}(1/N)\, ,
\label{eq:S-Laplace}
\end{align}
where as in Eqs.\ (\ref{eq:soft}) and (\ref{eq:WDYdef}) we identify $\omega=M(1-z)$, and define 
\bea
\bar{N} \equiv N e^{\gamma_E}\, ,
\label{eq:barNdef}
\eea
with $\gamma_E$ Euler's constant.\footnote{We note that the combination $\ln M/\bar N\mu$ is denoted by $L$ in \cite{Becher:2007ty}.}
 In the second equality of (\ref{eq:S-Laplace}) we have used  $z^N=e^{-(1-z)N}+{\cal O}(1-z)$.  Notice that the soft function, $\widetilde S$, is dimensionless in moment space and needs only two arguments.

\subsection{Evolution equation for the soft function}

The special role of the function $W_{a\bar a}(x_0\mu)$, Eq.\ (\ref{eq:softmatrixelement}) for threshold resummation was identified in this context by Korchemsky and Marchesini in Refs.\ \cite{Korchemsky:1992xv,Korchemsky:1993uz}.   It was evaluated at two loops 
by Belitsky \cite{Belitsky:1998tc}, who verified that in moment space it obeys a renormalization group equation
that generates double logarithms in moments,
\begin{align}
\label{eq:SNevolve}
\frac {d} {d \ln \mu} \ln \widetilde S \left(\ln \frac {M}{\bar N \mu}, \as(\mu) \right) &= -2\Gamma_{\rm cusp} \left(\alpha_s(\mu)\right) \ln \frac {M^2} {\bar{N}^2\mu^2}  -2  \gamma_W\left(\alpha_s(\mu) \right)\, ,
\end{align}
in terms of the cusp anomalous dimension, $\Gamma_{\rm cusp}(\as)$, and another anomalous dimension, $\gamma_W(\as)$ characteristic of this annihilation matrix element.\footnote{It is denoted $\gamma^W$ in Ref.\ \cite{Becher:2007ty}.}  Specifically, $\gamma_W$  in \cite{Becher:2007ty} equals $-(1/2)\Gamma_{\rm DY}$ in the notation of Belitsky \cite{Belitsky:1998tc}.
  This equation for the soft function summarizes the moment and momentum fraction evolution equations
in Refs.\ \cite{Sterman:1986aj} and \cite{Contopanagos:1996nh}, and reappears in this form in \cite{Becher:2007ty}, for
example.

The solution to the soft function evolution equation, Eq.\ (\ref{eq:SNevolve}), is
\begin{eqnarray}
\widetilde S\left(\ln \frac {M}{\bar N \mu_1},\as(\mu_1)\right) &=& \widetilde S\left( \ln \frac {M}{\bar N \mu_2},\as(\mu_2)\right)
\nn\\
&\ & \hspace{-10mm} \times\ 
\exp \left\{ \int_{\mu_2}^{\mu_1} \frac{d\mu'}{\mu'}\ \left(4 \Gamma_{\rm cusp} \, \ln\left( \frac{\mu' \bar N}{M}\right) -2 \gamma_W \left(\as(\mu')\right) \right) \right\}\, .
\label{eq:soln-gen}
\end{eqnarray}
It is only at this stage that SCET and dQCD methods part ways, in their choices of the  scales,
and then in their use of the inverse Laplace or Mellin transform to derive physical cross sections.

In the direct QCD resummation approach \cite{Contopanagos:1996nh,Belitsky:1998tc} the evolution equation \eqref{eq:SNevolve} is run from scale $\mu_2=M/\bar N$ to a factorization scale $\mu \sim M$, giving
\begin{align}
\widetilde S^\dq\left( \ln \frac {M}{\bar N \mu}, \as(\mu)\right) & 
\nn\\
& \hspace{-40mm} =\
S\left( 0,\as(M/\bar N)\right)\
\exp \left\{ \int_{M/\bar N}^\mu \frac{d\mu'}{\mu'}\ \left(4\Gamma_{\rm cusp} \, \ln\left( \frac{\mu' \bar N}{M}\right) -2 \gamma_W\left(\as(\mu')\right) \right) \right\}
\nn\\
& \hspace{-40mm}=\ \widetilde S\left( 0,\as(\mu)\right)\
\exp \left\{ \int_{M/\bar N}^\mu \frac{d\mu'}{\mu'}\ \left(4\Gamma_{\rm cusp}(\as(\mu')) \, \ln\left( \frac{\mu' \bar N}{M}\right) - \hat D\left( \as(\mu')\right)\ \right)\right\}\, .
\label{eq:S-soln-qcd-1} 
\end{align}
The first equality is the form that appears in Ref.\ \cite{Belitsky:1998tc}, and the second is a form given in \cite{Becher:2007ty} and \cite{Becher:2006mr} in comparing dQCD and SCET resummations.  In the second equality, the function $\hat D$ is defined by
\begin{align}
\hat D\left(\as(\mu')\right) = 2 \gamma_W\left(\as(\mu')\right)  + \mu' \frac{\partial }{\partial \mu'} \ln \widetilde S(0,\as(\mu')) \, .
\label{eq:hatD-def} 
\end{align}
The resummation function $\hat D(\as)$ is thus a hybrid object, the sum of an anomalous dimension and the logarithmic derivative of the non-local perturbative soft function.   We will return to its interpretation in the following section.

With a change of variables for the $\hat D$ term, we can rewrite Eq.\ (\ref{eq:S-soln-qcd-1}) as
\begin{align}
\label{eq:S-soln-qcd-2}
\widetilde S^\dq\left( \ln \frac{M}{N\mu},\as(\mu)\right) &= \widetilde S\left( 0,\as(M)\right)
\nn\\
& \hspace{5mm} \times\
\exp \left\{ \int_{1/\bar N}^{\mu/M} \frac{dy}{y} \left[\, 4 \int_{yM}^\mu \frac{d\mu'}{\mu'}\ \Gamma_{\rm cusp}(\as(\mu')) - \hat D\left( \as(yM )\right)\ \right] \right\}\, ,
\end{align}
which is a commonly used form, for example in \cite{Catani:2003zt} if we choose $\mu=\mu_f$.    We will discuss factorization scale dependence below.   Using explicit forms of the running coupling, Eq.\ (\ref{eq:S-soln-qcd-2}) may be evaluated analytically  to give the resummed moment dependence, as in Ref.\ \cite{Catani:2003zt}.   Any such expression will of course produce a Landau pole in the $N$ plane at $\bar N = M/\Lambda_{\rm QCD}$.   This singularity and its treatment in hadronic cross sections (see the next subsection) is sometimes cited as a motivation for the SCET treatment of resummation, to which we now turn.

For SCET, in Ref.\ \cite{Becher:2007ty} the running in Eq.\ (\ref{eq:SNevolve}) is taken directly between a short distance scale $\mu$
and a fixed, $N$-independent soft scale $\mu_s$,
\begin{align}
\widetilde S^\sc\left( \ln \frac{M}{\bar N \mu},\as(\mu),\mu_s\right) &= \widetilde S\left( \ln \frac{M}{\bar N\mu_s},\as(\mu_s)\right)
\nn\\
& \hspace{-25mm} \times\
\exp \left\{ \int_{\mu_s}^{\mu} \frac{d\mu'}{\mu'}\ \left(4 \Gamma_{\rm cusp}(\as(\mu')) \, \ln\left( \frac{\mu' \bar N}{M}\right) -2 \gamma_W\left(\as(\mu')\right) \right) \right\}\, .
\label{eq:soln-scet-1}
\end{align}
Here, $N$-dependence is only in the the soft function evaluated at scale $\mu_s$ and in the explicit factor of $\ln N$ that
multiplies the cusp anomalous dimension in the exponent.   Despite this alternative representation, it is clear that the two expressions,
Eq.\ (\ref{eq:S-soln-qcd-1}) and (\ref{eq:soln-scet-1}) are identically equal when taken at the same choice of scale $\mu$ and evaluated to all orders, in which
case $\widetilde S^\sc$ is independent of $\mu_s$.  Dependence on $\mu_s$ remains in $\widetilde S^\sc$, however, when it is evaluated to finite order, or summed to a fixed logarithmic order.
Thus, $\mu_s$ should be included as an argument of $\widetilde S^\sc$ in general.
Even in this case, however, the SCET and dQCD soft functions will be equal if we make the choice, $\mu_s=M/\bar N$.
We can summarize these results as
\bea
\widetilde S^\sc\left( \ln \frac{M}{\bar N\mu},\as(\mu)\right)_{{\rm all\ orders}}
&=&
\widetilde S^\dq\left( \ln \frac{M}{\bar{N\mu}},\as(\mu)\right)_{{\rm all\ orders}}\, ,
\nonumber\\
\widetilde S^\sc\left( \ln \frac{M}{\bar{N}\mu},\as(\mu),\mu_s=\frac{M}{\bar{N}}\right)_{{\rm fixed\ order}}
&=&
\widetilde S^\dq\left( \ln \frac{M}{\bar{N\mu}},\as(\mu)\right)_{{\rm fixed\ order}}\, .
\label{eq:equality}
\eea
 The free choice in the dQCD treatment that corresponds to the SCET soft scale is simply the lower limit of the $y$ integral in Eq.\ (\ref{eq:S-soln-qcd-2}), where $\bar N=e^{\gamma_E}N$ can be replaced by an arbitrary constant times $N$ at the same logarithmic accuracy, adjusting the function $\hat D$ at higher orders as necessary.

In Ref.\ \cite{Becher:2007ty}, the SCET expression (\ref{eq:soln-scet-1}) is further reorganized by using the $\ln N$ dependence in
the exponent as a generating function for $\ln N$-dependence in the soft function.   For this purpose, we adopt a notation close to that of Ref.\ \cite{Becher:2007ty} and write (\ref{eq:soln-scet-1}) as
\begin{align}
\label{eq:soln-scet-2}
\widetilde S^\sc \left(\ln \frac {M}{\bar{N\mu}},\as(\mu),\mu_s \right) =& \exp \left[ -4 S_{\rm cusp}\left(\mu_s,\mu \right) + 2 \alpha_{\gamma_W} \left(\mu_s, \mu \right) + \eta(\mu_s, \mu) \, \ln \frac {M^2}{ \mu_s^2}\right] 
\nn\\
& \hspace{5mm} \times \ \widetilde S \left(\ln \frac {M^2}{ \mu_s^2} +\frac{\partial}{\partial \eta(\mu)}, \mu_s \right)\
\exp\left[ -\ \eta(\mu_s, \mu) \, \ln \bar{N}^2 \right]
\, ,
\end{align}
where the various quantities in the exponents are defined by
\begin{align}
S_{\rm cusp}\left(\mu_s,\mu \right) &= -\int\limits_{\mu_s}^{\mu} \frac{d \mu'}{\mu'}  \, \Gamma_{\rm cusp}(\as(\mu')) \ln \frac {\mu'} {\mu_s}
\ =\ -\ \frac{1}{2}C_a\frac{\as}{\pi} \, \ln^2\frac{\mu}{\mu_s}\ \ +\cdots\,,
 \nonumber \\ 
 \alpha_{\gamma_W} \left(\mu_s,\mu \right) &= \ -\ \int\limits_{\mu_s}^\mu \frac{d \mu'}{\mu'}\, \gamma_W (\as(\mu'))
 \ =\ {\cal O}\left(\as^2\right) \, , 
 \label{eq:alphadef}
\end{align}
and
\begin{equation}
\eta(\mu_s, \mu) =\ -\ 2\alpha_{\Gamma} \left(\mu_s,\mu\right) = -2 \int\limits_{\mu_s}^{\mu} \frac{d \mu'}{\mu'} \, \Gamma_{\rm cusp}(\alpha(\mu'))
\ =\ -\ 2C_a\frac{\as}{\pi} \, \ln\frac{\mu}{\mu_s}\ +\ \cdots\, .
\label{eq:etadef}
\end{equation}
Here, for definiteness and later use, we show order $\as$ expressions, in which $a=q,g$, with $C_q=C_F$ and $C_g=C_A$.\footnote{In Ref.\ \cite{Becher:2007ty} the quantities in Eq. \eqref{eq:alphadef} and \eqref{eq:etadef} are evaluated by changing variables $d \ln \mu = d\alpha / \beta(\alpha)$, the details of which are omitted here.} Despite the differences in notation, every term in the resummed effective theory exponential has an exact correspondence in the direct QCD result, with differences due primarily to different choices of boundary conditions. Thus, for example, the combination $S_{\rm cusp}+\eta \ln (M^2/\mu_s^2)$ in (\ref{eq:soln-scet-2}) is precisely the $\Gamma_{\rm cusp}$ term of Eq.\ (\ref{eq:soln-scet-1}).  Matching to fixed-order and nonsingular calculations involving all parton types is also common to both formalisms, and we shall not discuss it here.

\subsection{Hadronic cross sections in direct QCD and SCET}

To close this section, we examine how the contrasting moment-space solutions (\ref{eq:S-soln-qcd-2}) and (\ref{eq:soln-scet-1}) are inverted and combined with parton distributions to produce physical cross sections, each with its individual estimate of higher order corrections.  In principle, both the moment-based dQCD expression (\ref{eq:S-soln-qcd-2}) and the SCET expression (\ref{eq:soln-scet-1}) can be transformed to give a hard-scattering function $C_{ab \rightarrow M} \left(z, M^2/\mu^2_f,\alpha_s(\mu_f)\right)$, Eq.\ (\ref{eq:refactorization}), directly in terms of the variable $z$, including all singular behavior for $z\rightarrow 1$, to an accuracy determined by the order to which the anomalous dimension $\Gamma_{\rm cusp}$ and the function $\hat D$, Eq.\ (\ref{eq:hatD-def}) are known \cite{Contopanagos:1993xh}.   The well-known ``minimal" approach \cite{Catani:1996yz} employed in Ref.\ \cite{Catani:2001ic,Catani:2003zt}, however, numerically inverts the product of the moment of the hard scattering function times the moment of the parton luminosity.
Before reviewing approaches to moment inversion, however, we will reintroduce dependence on an independent factorization scale, $\mu_f$ in the full cross section, Eq.\ (\ref{eq:basic-fact}).

The large-$N$ factorization scale dependence of the parton distributions is
\bea
\mu\frac{\partial}{\partial \mu} \ln f_{a/A} (N,\mu) &=&  -2A(\as(\mu)) \ln\bar N + B_a(\as(\mu)) + {\cal O}(1/N)
\nn\\
&=& -2 \Gamma_{\rm cusp}(\as(\mu)) \ln\bar N + 2\gamma_a(\as(\mu)) + {\cal O}(1/N)\, ,
\label{eq:fevol}
\eea
where the first form uses the conventional notation for DGLAP anomalous dimensions of parton distributions, and the second employs the relations $A=\Gamma_{\rm cusp}$ and  $B_a\equiv 2\gamma_a$ ($\gamma_a$ here is denoted by $\gamma^\phi$ in Ref.\ \cite{Becher:2007ty}).   
The solution at large $N$ for the distributions is then
\bea
f_{a/A} (N,\mu_f) &=& f_{a/A} (N,\mu)\ \exp \left[ \int_\mu^{\mu_f} \frac{d\mu'}{\mu'}\, \left( -2 \Gamma_{\rm cusp}(\as(\mu')) \ln\bar N + 2\gamma_a(\as(\mu') \right)\right]\, .
\label{eq:fmuf}
\eea
Substituting this relation into Eq.\ (\ref{eq:moment-product}) gives the factorized cross section with arbitrary factorization scale, simply by absorbing this exponential into the moments of the coefficient function $\widetilde C_{a\bar a}$, which near partonic threshold is given in momentum space by the refactorized form, Eq.\ (\ref{eq:refactorization}).  The Mellin moment of the inclusive cross section factorizes into products of moment-space parton distributions times hard-scattering functions, Eq.\ (\ref{eq:moment-product}).  Following \cite{Catani:1996yz}, one may invert the products of these functions, choosing a contour that passes between the logarithmic branch cuts at negative real $N$ and the running coupling branch cuts at large real $N$, intersecting the real axis at ${\rm Re}(N)=n_0>0$ within the specified range.   The presence of the right-hand branch cut leads in general to contributions with $\tau>1$ which, however, are exponentially suppressed in the hard scale.

 The  ``minimal" prescription \cite{Catani:1996yz} just described may be represented as
\bea
\frac{d\sigma^{(a)}_{AB\rightarrow M}(S,M^2)}{dM^2} &=&  \sigma^{a\bar a}_0(S,M^2)\  \sum_{a\leftrightarrow \bar a}   \
\int_{n_0-i\infty}^{n_0+i\infty} \frac{dN}{2\pi i}\, \tau^{-N}\  \widetilde {\cal L}_{a \bar a}(N,\mu_f)\
\nn\\
&\ & \hspace{20mm} \times\ 
 \widetilde C^\dq_{a\bar a}\left(N,\frac{M}{\mu_f},\as(\mu_f)\right)\, ,
\label{eq:qcd-min0}
\eea
where we exhibit only those partonic channels that require threshold resummation, and where from Eq. \eqref{eq:S-soln-qcd-2}, taking $\mu=M$ there, we have
\bea
 \widetilde C^\dq_{a\bar a}\left(N,\frac{M}{\mu_f},\as(\mu_f)\right) &=&  \hat  H_{a\bar a} \left(\frac{M}{\mu_f},\as(\mu_f) \right)\
\nn\\
&\ & \times
\exp\left\{ \int_{1/\bar N}^{1} \frac{dy}{y} \left[\, 4\int_{yM}^{\mu_f} \frac{dq}{q}\ \Gamma_{\rm cusp}(\as(q))\ -\ \hat D\left( \as(yM )\right)\ \right] \right\}
\nn\\
&\ & \hspace{20mm}  +\ {\cal O}(1/N)\, .
\label{eq:qcd-min}
\eea   
In this expression we have absorbed part of the  $\mu_f$ behavior into a modified hard scattering function, $\hat H_{a\bar a}$, which is simply the short distance function in Eq.\ (\ref{eq:refactorization}) multiplied by a factor associated with the evolution of the parton distributions,
\bea
\hat H_{a\bar a} \left(\frac{M}{\mu_f},\as(\mu_f) \right)
&=&
 \exp \left\{ -\ \int_M^{\mu_f} \frac{d\mu'}{\mu'}\; 4\gamma_a(\as(\mu'))\right\}\ H_{a\bar a} \left(1,\as(M) \right)\, .
\label{eq:HhatH}
\eea
The exponential factor absorbs all $\mu_f$ dependence associated with the anomalous dimensions $\gamma_a=\gamma_{\bar a}$.
The expansion of the transform of Eq.\ (\ref{eq:qcd-min}) will reproduce all leading-power singularities in $1-z$ to an order limited only by our knowledge of the anomalous dimension $\Gamma_{\rm cusp}$, the resummation function, $\hat D$ and the overall function $\hat H$.   Of course, to this expression we must add terms that are nonsingular for $z\rightarrow 1$ by matching.

The effective theory approach of Ref.\ \cite{Becher:2007ty} inverts moments of the soft function directly, before combining with the parton distribution functions.    The cross section at arbitrary factorization scale $\mu_f$  is just a restatement of Eq.\ (\ref{eq:moment-product}) with the SCET coefficient function,
\bea
\frac{d\sigma^{(a)}_{AB\rightarrow M}(S,M^2)}{dM^2} &=& 
\sigma^{a\bar a}_0(S,M^2)\ 
	     \sum_{a \leftrightarrow \bar a} 
	 \int_\tau^1 \frac{dz}{z}\ {\cal L}_{a \bar a}\left( \frac{\tau}{z},\mu_f \right)\ C^\sc_{a\bar a \rightarrow M} \left(z, M,\mu_f,\mu_s\right)\, .
	 \label{eq:sigma-scet}
\eea
In this case, the perturbative coefficient function is given at leading power in $1-z$ by combining the coefficient at scale $\mu$, Eq.\ (\ref{eq:soln-scet-2}) with the evolution factor Eq.\ (\ref{eq:fmuf}),
\bea
C^\sc_{a \bar a \rightarrow M} \left(z, M,\mu_f,\mu_s\right)
&=& \ H_{a\bar a}\left(\frac{M}{\mu},\as(\mu) \right)\ \int_{n_0-i\infty}^{n_0+i\infty} \frac{dN}{2\pi i}\ z^{-N}
\nn\\
&\ & \hspace{-40mm} \times\
\exp \left[ \int_\mu^{\mu_f} \frac{d\mu'}{\mu'}\, \left( -4 \Gamma_{\rm cusp}(\as(\mu')) \ln\bar N + 4\gamma_a(\as(\mu') \right)\right]\,
   \widetilde S_{a\bar a}^\sc \left( \frac{M}{\bar N\mu}, \as(\mu),\mu_s \right)\, .
   \nn\\
\label{eq:CNscet}
\eea
The result is formally independent of the starting factorization scale, $\mu$, which now plays the role of the ``hard [matching] scale" in the effective theory treatment of Ref.\ \cite{Becher:2007ty}.
When $N$ dependence in the boundary condition has been replaced by a series of derivatives, the inverse transform from $N$ to $1-z$ can be done explicitly, to give
 \begin{align}
\label{eq:softMomentum}
C_{a\bar a}^\sc \left( z, M,\mu,\mu_f,\mu_s \right) &= 
\nn\\
& \hspace{-35mm}H_{a\bar a}\left (\frac{M}{\mu}, \as(\mu)\right)
\exp \left[ -4 S_{\rm cusp}\left(\mu_s,\mu \right) + 2 \alpha_{\gamma_W} \left(\mu_s, \mu \right) 
+ 4\alpha_{\gamma_a}(\mu,\mu_f)+ \eta(\mu_s, \mu_f) \, \ln \frac {M^2}{ \mu_s^2}\right] 
\nn\\
&\hspace{-20mm} \times \ \widetilde S_{a\bar a} \left(\ln \frac {M^2}{ \mu_s^2} +\frac{\partial}{\partial \eta(\mu_s, \mu_f)}, \mu_s \right)\
 \frac {e^{-2\gamma_E \eta(\mu_s,\mu_f)}\, (1-z)^{2\eta(\mu_s,\mu_f)-1}} {\Gamma\left( 2\eta(\mu_s, \mu_f)\right ) } \ ,
\end{align}
where in  the exponential, the functions $S_{\rm cusp}(\mu_s,\mu)$ (not to be confused with $\widetilde S_{a\bar a}$), $\alpha_{\gamma_{\rm W}}(\mu,\mu_f)$ and $\eta(\mu_s,\mu_f)$ are given in Eqs.\ (\ref{eq:alphadef}) and (\ref{eq:etadef}).    The resummed cross section is then found directly in momentum space from Eq.\ (\ref{eq:sigma-scet}).    

In the treatment of the effective theory described in Ref.\ \cite{Becher:2007ty}, the soft scale $\mu_s$ is chosen to stabilize the hadronic cross section when $S_{a \bar a}$ is known to fixed order.   Notice that, in contrast to the dQCD result, Eq.\ (\ref{eq:qcd-min}), the leading powers in logarithms, $\as^k\ln^{2k-1}(1-z)$, are reproduced in the SCET result, Eq.\ (\ref{eq:softMomentum}), only to the order at which the soft function has been computed.   Beyond this order, leading logarithms in $1-z$ are replaced by combinations of logarithms of $1-z$ and $\mu_s$, as observed for example in \cite{Ahrens:2011mw}.   This need not be a problem, so long as the range $1-\mu_s/M<z<1$  is not phenomenologically important for the specific parton luminosity under consideration.  

To close this section, we summarize by comparing Eqs.\ (\ref{eq:qcd-min0}) for dQCD and (\ref{eq:sigma-scet}) for SCET.   When expanded in powers of $\as$, the dQCD minimal prescription reproduces leading and subleading logarithms in $1-z$ at all orders in perturbation theory, without introducing explicit nonperturbative parameters.    Of course the choice of Mellin inversion transform of Eq.\ (\ref{eq:qcd-min0}) that makes it possible to sum these terms is itself a particular choice of nonperturbative information.    The effective theory resummation  avoids the potential Landau pole in its application to hadronic cross sections.   It does so by replacing leading logarithms of $1-z$ beyond the lowest orders with a mixture of logarithms of $1-z$ and $\mu_s/M$, so that it remains strictly perturbative.   In Secs. \ref{sec:partonic} and \ref{sec:hadronic} we will explore the consequences of these rather different choices.   First, however, we relate our previous resummed exponents to exponents written in terms of inverse Mellin moments and plus distributions in the dQCD formalism.

\section{The soft function as an exponentiated Mellin moment}
\label{sec:webevolve}

We have rederived in Eq.\ (\ref{eq:qcd-min}) one of the basic dQCD forms for threshold resummation in correspondence to the effective theory treatment, but there is another form of the resummed cross section, which bears further comparison here.  In this form, the coefficient function $C_{a\bar a}$ of Eq.\ (\ref{eq:moment-product})  is given as the exponential of an explicit Mellin moment \cite{Sterman:1986aj,Catani:1989ne}, 
\bea
\label{eq:qcdmoment}
\widetilde C_{a \bar a}^\dq \left( N, \frac{M}{\mu_f}, \as\left(\mu_f\right) \right) &=&
\hat H_{a\bar a} \left(\frac{M}{\mu_f},\as(\mu_f) \right)\  \widetilde S_{a\bar a}\left (\frac{M}{\bar N\mu_f},\as(\mu_f)\right) 
\nonumber\\
&\ & \hspace{-20mm}=\  \widetilde H_{a\bar a} \left(\frac{M}{\mu_f},\as(\mu_f) \right)\ \exp \left \{ \int_0^1 dz \frac{z^{N-1}-1} {1-z} \right. 
\nn\\
&\ & \left. \hspace{-5mm} \times \left[ 4\int_{\mu_f}^{(1-z) M} \frac{d\mu'}{\mu'} A \left(\alpha\left(\mu'\right)\right) + 
D \left( \alpha\left( (1-z) M \right) \right) \right] \right \}\, ,
\eea
where the factorization scale dependence of the hard scattering function, $\hat H_{a\bar a}$ is given in Eq.\ (\ref{eq:HhatH}). In the second line we have changed $\hat H_{a\bar{a}}$ to $\widetilde H_{a\bar{a}}$ because, as we shall see below, part of the $N$-independent term in the soft function $S_{a\bar{a}}$ has been absorbed.  All leading, as well as many non-leading, logarithms of $N$ in Eq.\ (\ref{eq:qcdmoment}) are generated by the ``universal" anomalous dimension, $A(\as)=\Gamma_{\rm cusp}(\as)$, which is defined by the singular term, $2A(\as)/(1-x)_+$, in diagonal DGLAP evolution (parton $a$ here), or alternatively the coefficient of $\ln N$ in moment space, see Eq.\ (\ref{eq:fevol}).    The function, $D(\as)$ generates the remainder of non-leading logarithms of $N$.   It is clearly related to the function $\hat D(\as)$ in Eq.\ (\ref{eq:S-soln-qcd-1}), and we will rederive the rather complex expression for this relation, given in \cite{Catani:2003zt}.  We will also rederive Eq.\ (\ref{eq:qcdmoment}) below from our previous considerations.   This will enable us to give a direct interpretation of $D(\as)$ in terms of the soft function, $S_{a\bar a}$ in $z$ space that is as natural as the definition of $\hat D(\as)$ in Eq.\ (\ref{eq:hatD-def}), which is formulated in terms of the soft function, $\widetilde S_{a\bar a}$ in $N$-space.

\subsection{Evolution for the exponent}

To derive Eq.\ (\ref{eq:qcdmoment}), we consider the logarithm of the moment space soft function,
\bea
\widetilde E  \left( \ln \frac{M}{\bar N\mu},\as(\mu)\right)\ \equiv\ \ln \left[\ \widetilde S\left( \ln \frac{M}{\bar N\mu},\as(\mu)\right) \ \right]   \, .
\label{eq:Edef}
\eea
The function $\widetilde E$ can be defined as the sum of a set of modified perturbative diagrams, so-called
``webs" \cite{Sterman:1981jc,Gatheral:1983cz,Frenkel:1984pz}, which are discussed in the context of resummation  in Ref.\ \cite{Laenen:2000ij}.   We will not reproduce their explicit construction here, but only emphasize that they provide a well-defined perturbative expansion that begins at order $\as$ with single gluon exchange and emission.

The inverse transform of $\widetilde E$ to momentum space is
\begin{align}
E \left(1-z,\frac{M(1-z)}{\mu}, \alpha(\mu)\right) 
&= \int_{n_0-i\infty}^{n_0+i\infty} \frac{dN}{2\pi i}\, z^{-N}\  
\widetilde E \left( \ln \frac {M}{\bar N \mu}, \alpha\left(\mu\right) \right) 
 \nonumber \\
 & \hspace{-20mm} = \int_{n_0-i\infty}^{n_0+i\infty} \frac{dN}{2\pi i}\, e^{N(1-z)}\  
\widetilde E \left( \ln \frac {M}{\bar N \mu}, \alpha\left(\mu\right) \right) \ +\ {\cal O}\left((1-z)^0\right)\, .
\label{eq:E-Mellin-Laplace}
\end{align}
Applying  the inverse Laplace transform to the evolution equation (\ref{eq:SNevolve}) for $\widetilde E=\ln \widetilde S$, we also derive an evolution equation for the same momentum space the web function, $E$, accurate to leading power in $1-z$, 
\bea
\label{eq:webEvolve}
\frac{d}{d\ln\mu} E \left(1-z, \frac{M(1-z)}{\mu}, \alpha(\mu) \right) &=& \left[ -4\Gamma_{\rm cusp} \left(\alpha(\mu)\right) \ln \frac {M} {\mu} -2\gamma_W \left(\alpha(\mu) \right) \right] \delta(1-z) \nonumber \\
&\ &  \quad -\ 4 \Gamma_{\rm cusp} \left(\alpha(\mu)\right) \left( \frac{1}{1-z} \right)_+.
\eea
A general solution to this equation for $z<1$ is
\bea
\label{eq:E-z-soln}
 E \left(1-z, \frac{M(1-z)}{\mu}, \alpha(\mu) \right) &=& 
 E \left(1-z, 1, \alpha_s((1-z)M) \right)\ 
 \nonumber\\
 &\ & \hspace{0mm}
 -\ \frac{4}{1-z}\; \int_{(1-z)M}^{\mu} \frac{d\mu'}{\mu'}\, \Gamma_{\rm cusp}\left(\as(\mu')\right)\,  ,
 \eea
where we have used the natural scale $(1-z)M$ as the starting point of evolution. Setting $\mu=M$ and taking the Mellin moment of $E(1-z)$, we derive precisely the form of the exponent in Eq.\ (\ref{eq:qcdmoment}) up to a constant that multiplies $\delta(1-z)$  (and to which we will return in the next subsection), 
\bea
\label{eq:Dz-to-E}
\frac{D \left( \alpha\left( (1-z) M \right) \right)}{1-z} &=&\ E \left(1-z, 1, \alpha_s((1-z)M) \right) \ {\rm for} \ 1-z>0\, .
\eea
We can confirm the overall factor $1/(1-z)$ in the web function, $E$  by a simple dimensional analysis of the second equality of Eq.\ \eqref{eq:E-Mellin-Laplace}. Comparing Eqs.\ \eqref{eq:Dz-to-E} and (\ref{eq:E-z-soln}), we recognize $D(\as)/(1-z)$ as the remainder of the web function when the soft function is collinear-subtracted (or equivalently, UV renormalized \cite{Belitsky:1998tc}) at scale $(1-z)M$.
This definition of the $D$ term in Eq. \eqref{eq:Dz-to-E}  reproduces the Drell-Yan $D^{(2)}$ coefficient \cite{Vogt:2000ci} from the two-loop result for the soft function in \cite{Belitsky:1998tc}.

We note that Eq.\ (\ref{eq:Dz-to-E}) for function $D(\as)$ is comparable in simplicity to the definition of $\hat D$ in Eq.\ (\ref{eq:hatD-def}), and gives a transparent interpretation of $D$ in terms of the non-local soft function.
This is to be contrasted to the relative complexity of the relationship between $D$ and $\hat D$, found by comparing Eqs.\ (\ref{eq:qcdmoment}) and (\ref{eq:S-soln-qcd-1}),  \cite{Catani:2003zt}
\begin{align}
\hat D (\alpha_s) &= e^{2 \gamma_E \nabla} \Gamma \left( 1+2\nabla \right)  D\left( \alpha_s\right) + \frac {e^{2 \gamma_E \nabla} \Gamma \left( 1+2\nabla \right) - 1}{\nabla} 2A \left( \alpha_s \right), 
\label{eq:relationD} 
\end{align}
where following the notation of \cite{Becher:2007ty} we define
\begin{eqnarray}
\nabla &\equiv \frac{d}{d \ln \mu^2}\ \equiv \frac{\beta \left(\as(\mu)\right)}{2}\, \frac{\partial}{\partial \as(\mu)}\, .
\end{eqnarray}
The  action of the derivatives $\nabla$ is interpreted through a Taylor expansion, keeping in mind that each derivative acts only on the running coupling and therefore promotes the order in $\as$.   For completeness, we re-derive Eq.\ (\ref{eq:relationD}) in Appendix \ref{sec:appendix}.  

 From Eq.\ \eqref{eq:relationD} in combination with the expression for $\hat D$ in Eq.\ \eqref{eq:hatD-def} we obtain the differential identity found in \cite{Becher:2007ty} relating the $D$ function to the anomalous dimension $\gamma_W$ and the soft function in moment space,
 \begin{align}
e^{2 \gamma_E \nabla} \Gamma \left( 1+2\nabla \right) \frac {D\left( \alpha_s\right)} {2} &=  \gamma_W \left( \alpha_s\right) + \nabla \ln \widetilde S \left(0,\alpha_s \right) - \frac {e^{2 \gamma_E \nabla} \Gamma \left( 1+2\nabla \right) - 1}{\nabla} A \left( \alpha_s \right)\, .
\label{eq:relationD1}
\end{align}
The complexity of this expression compared to Eq.\ (\ref{eq:Dz-to-E}) is precisely due to using the moment-space soft function $\widetilde S$ (evaluated at $\mu=M/\bar N$) on the right-hand side in an expression for the momentum space function $D(\as)$ on the left.

\subsection {Order-by-order structure of the exponent}

We can gain further insight into the $D$ function and its relationship to the soft function non-leading anomalous dimension, $\gamma_W$ in Eq.\ (\ref{eq:SNevolve}), by studying the order-by-order expansion of the full exponent, $E$, Eq.\ (\ref{eq:Edef}).
We start by expanding $E$ at  $\mu=M$ \cite{Eynck:2003fn, Laenen:2005uz},
\begin{align}
E(1-z,1-z,\as(M)) &= 
\int_{n_0-i\infty}^{n_0+i\infty} \frac{dN}{2\pi i}\, e^{N(1-z)}\  
\widetilde E \left( \ln \frac {1}{\bar N }, \alpha_s\left(M\right) \right) \ 
 \nonumber \\
&=\ F\left(\alpha_s\left(M\right) \right) \delta(1-z) + D\left(\alpha_s\left(M\right) \right) \, \left( \frac {1} {1-z} \right)_+ 
\nonumber \\ &
\hspace{20mm} +\ \sum\limits_{k=1}^\infty E^{(k)}\left(\alpha_s\left(M\right) \right) \, \left( \frac {\ln^k (1-z)}{1-z} \right)_+\, .
\label{eq:webM}
\end{align}
We will see that  the function $D(\as)$ here  turns out to be the same function as above. 
The expansion for general choice of scale $\mu$ is found from a change of variable in the inverse transform,
\begin{align}
E\left(1-z,\frac{(1-z)M}{\mu},\as(\mu)\right) 
&=  \int_{n_0-i\infty}^{n_0+i\infty} \frac{dN}{2\pi i}\, e^{N(1-z)}\  
\widetilde E \left( \ln \frac {M}{\bar N \mu}, \alpha_s\left(\mu\right) \right) \ 
\nn \\
&= \frac{M}{\mu}\; \int_{n_0-i\infty}^{n_0+i\infty} \frac{dN'}{2\pi i}\, e^{N'\frac{M(1-z)}{\mu}}\  
\widetilde E \left( \ln \frac {1}{\bar N' }, \alpha_s\left(\mu\right) \right) 
\nn \\ 
&= \frac{M}{\mu}\; \int_{n_0-i\infty}^{n_0+i\infty} \frac{dN'}{2\pi i}\, e^{N'(1-z')}\  
\widetilde E \left( \ln \frac {1}{\bar N' }, \alpha_s\left(\mu\right) \right),
 \label{eq:webMu1}
\end{align}
where in the third equality we have defined
\bea
1-z' \equiv \frac{(1-z)M}{\mu}\, .
\label{eq:1-z'def}
\eea
The inverse transform in final line of (\ref{eq:webMu1}) has the same form as in (\ref{eq:webM}), but now the distributions in its expansion are in terms of $1-z'$ rather than $1-z$.  In particular, the resulting plus distributions are defined to give zero when integrated from $z'=0$ to $z'=1$, rather than from $z=0$ to $z=1$.    The relations between the two sets of distributions are, however, simple,
\bea
\delta(1-z') &=& \frac{\mu}{M}\, \delta(1-z)\, ,
\nn\\
\left(\frac {\ln^k (1-z')} {1-z'} \right)_+
&=&
\frac{\mu}{M}\,  \left[\ \left(\frac {\ln^k M(1-z)/\mu} {1-z} \right)_+ + \frac{\ln^{k+1} M/\mu}{k+1} \delta(1-z)\ \right]\, .
\eea
In these terms, the web function, Eq.\ \eqref{eq:webMu1} is equal to $\frac M \mu$ times Eq. \eqref{eq:webM}, with $1-z$ replaced by $1-z'$,
\begin{align}
\label{eq:webMu}
E\left(1-z,\frac{(1-z)M}{\mu},\as(\mu)\right)  &=
\nn\\
& \hspace{-20mm}
 \left[ F(\alpha_s(\mu)) + D(\alpha_s(\mu)) \ln \frac M \mu + \sum\limits_{k=1}^\infty E^{(k)}(\alpha_s(\mu)) \frac {\ln^{k+1} M/\mu}{k+1} \right] \delta (1-z) \nonumber \\
& \hspace{-20mm} + D(\alpha_s(\mu)) \, \left( \frac {1} {1-z} \right)_+ + \sum\limits_{k=1}^\infty E^{(k)}(\alpha_s(\mu)) \, \left( \frac {\ln^k M(1-z)/\mu }{1-z} \right)_+\, .
\end{align}
Now by imposing $\mu=M(1-z)$ in \eqref{eq:webMu} for $z\ne 1$ we rederive Eq.\ (\ref{eq:Dz-to-E}) in the form
\bea
\label{eq:Drelation}
E\left(1-z,1,\as((1-z)M)\right)  = \frac{1}{1-z}\ D \left(\alpha_s(\mu)\right)\, .
\label{eq:Dzne1}
\eea
By demanding that the $\mu$ derivative of the delta function terms in the general expansion, (\ref{eq:webMu}), coincide with the delta function part of the evolution equation for the exponent (\ref{eq:webEvolve}), with $\mu=M$, we immediately find as well an expression for $\gamma_W(\as)$,
\bea
\gamma_W(\as(\mu)) = \ -\ \frac{1}{2}\ \left[\, \mu \frac{d}{d\mu}\, F(\as(\mu)) \ -\ D(\as(\mu))\, \right]\, ,
\label{eq:GDY-FD}
\eea
in terms of the $D$ function and the $z$-independent terms in the expansion of the exponent, Eq.\ \eqref{eq:webMu}.  Again, the simplicity of this relation relative to  Eq.\ \eqref{eq:relationD1} results from working consistently in $z$ space. This relation is in fact the same as Eq. (4.4) in Ref. \cite{Laenen:2005uz}, if we apply the relation Eq. (44) in \cite{Becher:2007ty} and make the assumption
\bea
\label{eq:laenen-becher}
\tilde{G} = -\gamma^V.
\eea
By comparing explicit coefficients, we checked that Eq. \eqref{eq:laenen-becher} holds true up to 3-loop order.

\section{Logarithmic accuracy in the partonic cross section}
\label{sec:partonic}

The partonic cross section is a special case of the hadronic cross section, with parton distributions replaced by delta functions.
The relationship between dQCD and SCET coefficient functions has been studied  in both \cite{Becher:2007ty} and \cite{Bonvini:2012az}, where a pattern of agreement was demonstrated, when the soft scale $\mu_s$ is chosen as $M/\bar N$ in moment space.  Certainly, the moment space equality is already implicit in the comparison of the dQCD and SCET soft function, Eqs.\ \eqref{eq:S-soln-qcd-1} and \eqref{eq:soln-scet-1}, respectively.  It will be instructive, however, to follow the lead of \cite{Bonvini:2012az} as applied to the coefficient functions and partonic cross sections, to illustrate how the soft function evolution equation Eq. \eqref{eq:SNevolve} organizes the relationship between the SCET and dQCD coefficients for all choices of scales.  In this section we hope to clarify further this agreement at the partonic level, relying on the evolution equation for the exponential function, Eq.\ (\ref{eq:webEvolve}).   As in Ref.\ \cite{Bonvini:2012az}, we shall choose $\mu_f=M=\mu$, where our $\mu$ corresponds to the SCET hard matching scale, $\mu_h$ in Ref.\ \cite{Becher:2007ty}.

From the refactorized expression, Eq.\ (\ref{eq:refactorization}) and the soft function (\ref{eq:soln-scet-1}), the hard scattering function in the effective theory treatment can be written in terms of the single hard scale, $M$, and the (so far arbitrary) soft scale $\mu_s$, as
\begin{align}
\label{eq:scetMom}
\widetilde{C}_{a\bar a}^\sc \left( N, M, \mu_s \right) &=  H_{a\bar a} \left(\as(M) \right)\ 
\widetilde S_{a\bar a} \left(\ln \frac {M}{\bar N \mu_s}, \mu_s \right) \nonumber \\
& \quad \times\ \exp \int_{\mu_s}^{M} \frac {d\mu'{}}{\mu'{}} \left[ -4 \Gamma_{\rm cusp} \left( \alpha(\mu') \right) \ln \frac {M}{\bar N \mu} -2 \gamma_W \left( \alpha(\mu') \right)\right].
\end{align}
As noted in Sec.\ \ref{sec:soft}, $\widetilde C_{a\bar a}^\sc$ is in principle independent of $\mu_s$ at all orders, but since we are interested precisely in the effect of taking only finite orders in $\widetilde S_{a\bar a}$, we introduce the soft scale as an additional argument.   
Also, from Sec.\ \ref{sec:soft}, we may assume for this discussion that 
\begin{equation}
\label{eq:qcd-scet-N}
\widetilde{C}_{a\bar a}^\sc \left( N, M, \mu_s=M/\bar N \right) = \widetilde{C}_{a\bar a}^\dq \left( N, M \right)\, .
\end{equation}
This follows from Eq.\ (\ref{eq:equality}), assuming that the hard-scattering functions $H_{a\bar a}$ are handled identically. 
 
In any case, once the hard-scattering functions are treated equivalently, Eq.\ (\ref{eq:qcd-scet-N}) holds up to $1/N$ corrections when the two sides are evaluated to all orders, and up to subleading  logarithmic  corrections when the two sides are truncated to finite order. However, Eq. \eqref{eq:qcd-scet-N} does not immediately imply that partonic cross sections from the two methods agree, given that the transformations back to $z$ space are handled differently.  To demonstrate  their agreement, we will follow the method of \cite{Bonvini:2012az,Bonvini:2013td}, which investigates the difference between SCET and QCD predictions in terms of the ratio of their resummed moments. 

The ratio of the QCD moment to SCET moment for arbitrary $\mu_s$ is \cite{Bonvini:2012az}
\begin{align}
\label{eq:Cr}
\widetilde C_{r,(a\bar a)} \left( \ln \frac {M}{\bar N \mu_s},\as(\mu_s) \right)\ &=\ \frac {\widetilde{C}_{a\bar a}^\dq \left( N, M \right)} {\widetilde{C}_{a\bar a}^\sc \left( N, M, \mu_s \right)} \nonumber \\
&=\ \frac {\widetilde{C}_{a\bar a}^\sc \left( N, M, \mu_s=M/\bar N \right)} {\widetilde{C}_{a\bar a}^\sc \left( N, M, \mu_s \right)}\, ,
\end{align}
where if calculated to all orders, $C_{r,(a\bar a)}$ would equal $1$ identically.   The reorganization of the effective theory, however, breaks this identity, and we would like to know at what logarithmic order this begins.  

Subsitituting Eq. \eqref{eq:scetMom} into Eq. \eqref{eq:Cr} allows us to derive the explicit form, 
\begin{align}
\label{eq:Cr1}
\widetilde{C}_{r,{(a\bar a)}} \left(\ln \frac{M}{\bar N \mu_s}, \as (\mu_s) \right) = \exp \int_{M/\bar N}^{\mu_s} \frac {d\mu}{\mu} & \left\{ -4 \Gamma_{\rm cusp} \left( \alpha(\mu) \right)\ln \frac {M} {\bar N \mu} -2 \gamma_W \left( \alpha(\mu) \right) \right. \nonumber \\
& \left. \hspace{10mm}
 - \mu \frac{d}{d\mu}\ \ln \left[ \widetilde S_{a\bar a} \left(\ln \frac {M^2}{\bar N^2 \mu^2}, \mu \right) \right] \right\}\, ,
\end{align}
an expression that manifestly reduces to unity when $\mu_s = M / \bar N$.   We have promoted the ratio of soft functions evaluated at scales $\mu_s$ and $M/\bar N$ into an integral of their logarithmic derivative between these values, in the spirit of Eq.\ (\ref{eq:S-soln-qcd-1}) above.  Eq.\ (\ref{eq:Cr1}) generalizes the ``master formula" of Ref.\ \cite{Bonvini:2012az} beyond NNLL.   Now, in Eq.\ \eqref{eq:Cr1}, the integrand in the exponent vanishes when $\tilde S$, $\Gamma_{\rm cusp}$ and $\gamma_W$ are evaluated to all orders, because of the evolution equation for the soft function in moment space, Eq.\ \eqref{eq:SNevolve}.
However, for ${\rm N}^k$LL resummation in the effective theory treatment of  Ref.\ \cite{Becher:2007ty}, $\Gamma_{\rm cusp}$, $\gamma_W$, and $\widetilde S_{a\bar a}$ are truncated at order $\as^{k+1}$, $\as^k$ and $\as^{k-1}$, respectively.   In this case, the cancellation in the integrand for general $\mu_s$ is only exact up to order $\alpha_s^{k-1}$, and the exponent becomes non-zero at order $\as^k$, including leading logarithmms, $\as^k\ln^{2k}N$. This corresponds to the observation in \cite{Bonvini:2012az} that at NNLL, $C_r-1$ starts at order $\as^2$.  For ${\rm N}^k$LL resummation in the  convention for dQCD resummation described in \cite{Bonvini:2012az}, we would want to use $\widetilde S_{a\bar a}$ to one higher order, $\as^k$.  In this case the moment ratio $C_r$ deviates from unity  beginning at order $\as^{k+1}$, including leading terms that behave as $\as^{k+1}\ln^{2(k+1)}N$. This corresponds to the observation in \cite{Bonvini:2013td} that $C_r - 1$ starts at $\as^3$ order for NNLL resummation in the dQCD convention.
    
It is now instructive to revisit the method that leads to the
effective theory coefficient function, Eq.\ (\ref{eq:softMomentum}), using the ratio $C_r$ defined in Eq.\ (\ref{eq:Cr}).   We first rewrite the SCET result slightly, moving explicit $(1-z)$-dependence to the left of the soft function $\widetilde S_{a \bar a}$, Eq.\ (\ref{eq:soln-scet-2}), which leads to
\begin{align}
C_{a\bar a}^\sc \left( z, M, \mu_s \right) &= 
H_{a\bar a}\left (\as(M) \right)
\exp \left[ -4 S_{\rm cusp}\left(\mu_s,M \right) + 2 \alpha_{\gamma_W} \left(\mu_s, M \right) \right]
\nn\\
& \hspace{5mm} \times \frac {1} {(1-z)^{1-2\eta(\mu_s, M)}} \left( \frac{M^2}{\mu_s^2} \right)^{\eta(\mu_s, M)} \nonumber \\
   & \hspace{5mm} \times \widetilde S_{a \bar a} \left(\ln \frac{M^2(1-z)^2}{\mu_s^2} + \partial_{\eta(\mu_s,M)},\alpha\left(\mu_s\right) \right)\, 
  \frac {e^{-2\gamma_E \eta(\mu_s,M)}\, } {\Gamma\left( 2\eta(\mu_s,M)\right ) }\, . 
   \label{eq:scetKer}
\end{align}
Now, by Eq.\ (\ref{eq:Cr}), the QCD coefficient function is of exactly the same form as the SCET coefficient, Eq. \eqref{eq:scetMom}, but with $\widetilde S_{a \bar a}$ replaced by $\widetilde C_r \times \widetilde S_{a \bar a}$, where both $\widetilde C_r$ and $\tilde S_{a\bar a}$ depend on the moment variable only through the ratio $N\mu_s/M$.   As a result, the direct QCD
coefficient function
can be written as
\begin{align}
\label{eq:QCDscetKer}
C_{a\bar a}^\dq (z, M) &=
H_{a\bar a}\left(\as(M)\right) 
\exp \left[ -4 S_{\rm cusp}\left(\mu_s,M \right) + 2 \alpha_{\gamma_W} \left(\mu_s, M \right) \right]
\nn\\
& \hspace{5mm} \times \frac {1} {(1-z)^{1-2\eta(\mu_s,M)}} \left( \frac{M^2}{\mu_s^2} \right)^{\eta(\mu_s,M)} \widetilde C_{r,(a\bar a)} \left(\ln \frac{M^2(1-z)^2}{\mu_s^2} + \partial_\eta,\alpha\left(\mu_s\right) \right)
\nn\\
&\hspace{5mm}  \times\ \widetilde S_{a \bar a} \left(\ln \frac{M^2(1-z)^2}{\mu_s^2} + \partial_\eta,\alpha\left(\mu_s\right) \right)\,
  \frac {e^{-2\gamma_E \eta(\mu_s,M)}\, } {\Gamma\left( 2\eta(\mu_s,M)\right ) }\, . 
\end{align}
For the partonic cross section, we 
can
set $\mu_s = M(1-z)$ to eliminate logarithms in Eq.\ \eqref{eq:scetKer}. With this scale setting, the large logarithm appearing in $\widetilde C_r$ in Eq. \eqref{eq:QCDscetKer} is 
also eliminated. Since $\tilde S_{a\bar a}$ is evaluated to $\alpha^{k-1} \left(\mu_s \right)$, while $\widetilde C_r$ only deviates from $1$ starting at order $\alpha^k\left(\mu_s\right)$, we conclude that with the choice $\mu_s=M(1-z)$, the SCET partonic cross section agrees with the QCD partonic cross section up to subleading logarithms in moment space. 

The correspondence between dQCD and the SCET formalism in momentum space when $\mu_s=M/\bar N$ to next-to-next-to-leading logarithm is worked out explicitly in \cite{Bonvini:2012az}. 
Even at leading logarithms, however, it is useful to illustrate the difference between the resummed results at arbitrary $\mu_s$, and their agreement for $\mu_s=M(1-z)$.   At fixed coupling and leading logarithm, the SCET resummed 
coefficient function is found by isolating the leading terms of the general form, Eq.\ (\ref{eq:softMomentum}), and recalling the ${\cal O}(\as)$ results of Eqs.\ (\ref{eq:alphadef}) and (\ref{eq:etadef}), 
\begin{align}
C_{a\bar a}^{[SCET]} (z,M,\mu_s) &= \delta(1-z) +  4C_a\, \frac{\as}{\pi} \frac{\ln (\mu_s/M)}{1-z} \nonumber \\
 & \qquad \times \exp \left[2C_a\, \frac{\as}{\pi} \ln \frac{\mu_s}{M} \left(2\ln(1-z)-\ln \frac{\mu_s}{M} \right) \right]+\rm{NLL}\, .
\label{eq:scetLL}
\end{align}
For comparison, the dQCD coefficient in $z$ space can be found directly by the inverse moment of Eq.\ (\ref{eq:qcd-min}), keeping only the lowest-order term from $\Gamma_{\rm cusp}$ (the case $k=0$), to get
\begin{align}
C_{a\bar a}^{[dQCD]} (z,M) &= \delta(1-z) +
4C_a\, \frac{\as}{\pi} \frac{\ln (1-z)}{1-z} \exp \left[2C_a\, \frac{\as}{\pi} \ln^2 (1-z) \right] + \rm{NLL}\, .
\label{eq:qcdLL}
\end{align}
As expected, these two expressions coincide exactly for $\mu_s = M(1-z)$, and as we have observed, Eq.\ (\ref{eq:Cr1}) shows that this result extends to arbitrary logarithmic order, when more orders are included in the exponent.  In the SCET  LL example of Eq.\ \eqref{eq:scetLL}, for fixed values of $\mu_s$, however, the ${\cal O}(\as)$ singular behavior $[\ln(1-z)/(1-z)]_+$ is absent, and the two expressions agree only at zeroth order.   This illustrates the general pattern, that for ${\rm N}^k$LL resummation with $k \geq 1$, the SCET resummed coefficient contains explicit leading logs of $1-z$ up to order $\alpha^{k-1}$, while at higher orders leading logarithms of $1-z$ are replaced by monomials in logs of $\mu_s / M$ and $1-z$.

In summary, at the specific choice $\mu_s=M(1-z)$,
 the effective theory and direct QCD resummations can be regarded as essentially equivalent at the partonic level.  
 When $\mu_s$ is taken as a fixed quantity for hadronic cross sections, however,
 the situation is more complex.   We turn to this comparison in the next section. 

\section{Comparing resummed hadronic cross sections}
\label{sec:hadronic}

In this section we explore the relationship between dQCD and SCET hadronic cross sections, and show that their difference can be quantified in terms of an expansion of the partonic luminosity function, ${\cal L}_{a\bar a}(\tau/z)$, Eq.\ (\ref{eq:basic-fact}).  The leading term in this expansion turns out to be identical in dQCD and SCET for a natural choice of the soft scale, $\mu_s$,  up to other choices involving non-threshold corrections.   We will see that in many practical cases corrections beyond the leading term are small,  further subleading both in order of $\as$ and of logarithms.  This can be the case for nearly the full range of the variable $\tau$.\footnote{Reference \cite{Bauer:2010jv} also studies the relation of parton distribution shapes to threshold resummation within an SCET formalism.}
    
It is already clear from Eq.\ (\ref{eq:basic-fact}) that,  if the parton luminosity function ${\cal L}_{a\bar a}(\tau/z)$ is given by a single power of $\tau/z$, the hadronic cross section is approximately proportional to a simple moment of the hard scattering function \cite{Becher:2007ty} (see also \cite{Bonvini:2012an} for a similar conclusion derived from the saddle point approximation for the Mellin inversion integral).   In this approximation, we know from Eq.\ (\ref{eq:equality}) that the hadronic cross sections of the dQCD and SCET formalisms will be equal if the SCET soft scale is taken as $\mu_s=M/\bar s_1$,  with $s_1$ the effective power.  Thus, if such a criterion is adopted to determine $\mu_s$, the resummed cross sections will agree exactly. In practice, of course, this ``single power approximation'' is subject to finite corrections. We will show that such corrections are rather small for a wide range of $\tau$, and can be incorporated systematically into both dQCD and SCET formalisms. Indeed, we will also see that these corrections produce differences between dQCD and SCET resummed hadronic cross sections only at subleading logarithmic order when $\mu_s$ is chosen as above.  Note that this agreement will extend beyond the resummed expression to the fully-matched cross section, because the fixed-order moments in the two formalisms will agree automatically to the level that matching has been carried out.   

In the following, we first describe the consequences and test the validity of the single power approximation.   We go on to describe the expansion for partonic luminosity in which the single-power approximation is the leading term.  We will conclude with some tests of the expansion in realistic cases, comparing the SCET and dQCD formalisms in the process.

\subsection{The single-power approximation}
\label{subsec:single}

\begin{figure}
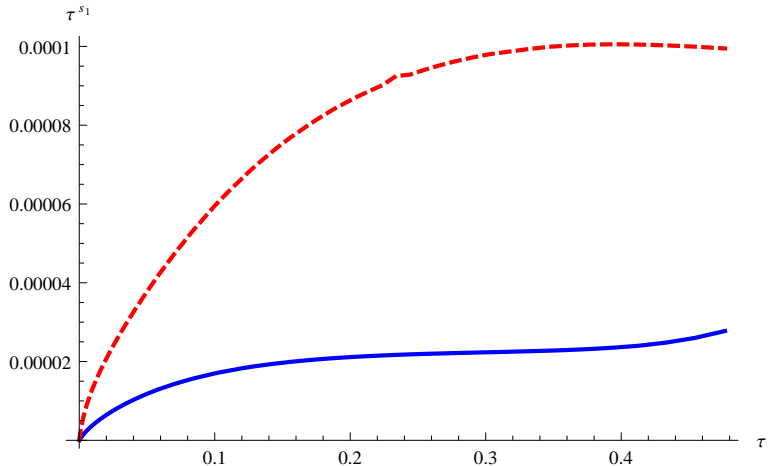

\centerline{\figscale{Tau_raised_to_s1_power}{10cm}}
\caption{Plots of correction factors $\tau^{s_1}$ in Eq.\ (\ref{eq:convol-moment}) for gluon-gluon luminosities from Ref.\ \cite{Martin:2009iq}.      The solid line represents factorization scale  126 GeV and dashed line 10 GeV. \label{fig:tau-singlepower}}
\end{figure}

Let us imagine that the parton luminosity really is exactly power-behaved.  This is our ``single-power approximation",
for parton $a$,
\bea
{\cal L}_{a\bar a}\left(\frac{\tau}{z}\right)\ &=&\ {\rm const}\ \left[\frac{\tau}{z}\right]^{-s_1(\tau)}\ =\ {\cal L}_{a\bar a}(\tau)\, z^{s_1(\tau)}\, ,
\label{eq:L-power}
\eea
where $s_1$ depends in general on the parton flavor $a$.   For the remainder of this section, we drop partonic indices.  As indicated, parameter $s_1$ is  a function of $\tau$ (and also of the factorization scale, which we suppress).
For such a luminosity, ${\cal L}(\tau/z)$, convolution with the corresponding partonic hard scattering coefficient $C(z)$ in Eqs.\ \eqref{eq:basic-fact} and \eqref{eq:moment-product} can be identified with the $s_1$-Mellin moment of $C(z)$,
\bea
\frac{d\sigma(\tau)}{dM^2} &=& \sigma_0\ \int_{\tau}^1 \frac{dz}{z} {\cal L}\left(\frac \tau z\right) C(z) \nn \\
&=& \sigma_0\ {\cal L}(\tau)\ \int_{\tau}^1 \frac{dz}{z} z^{s_1(\tau)}\, C(z) \label{eq:convol-moment-2}\\
&=& \sigma_0\ {\cal L}(\tau)\  \widetilde C \left(s_1(\tau)\right)\ +\ {\cal O}\left( \tau^{s_1} \right)\, .
\label{eq:convol-moment}
\eea
In the final equality we have assumed
\begin{equation}
1/s_1 \ll \ln\frac{1}{\tau}\, ,
\label{eq:tau-zero}
\end{equation}
which implies that for $z\sim \tau$, the integrand is suppressed by a factor $\tau^{s_1} \ll 1$.   This ensures that the effect of the nonzero lower limit $z>\tau$ in Eq.\ \eqref{eq:convol-moment} is negligible, so that \eqref{eq:convol-moment} becomes the full Mellin moment.   The property $\tau^{s_1}\ll 1$ is illustrated for a realistic gluon-gluon luminosity (which will be from the MSTW 2008 NNLO gluon distribution \cite{Martin:2009iq} throughout the rest of this paper) in Fig.\ \ref{fig:tau-singlepower}, with $s_1$ the logarithmic derivative $s_1\equiv d\ln{\cal L}(y)/d\ln y$ at $y=\tau$.  It can also be verified for almost the entire range of $\tau$ for realistic $q\bar q$ luminosities.   
Note that it does not by itself require $s_1$ to be a large number if $\tau$ is small.  Correspondingly, it is natural to expect that $s_1$ grows as $\tau$ increases.

The argument leading to Eq.\ (\ref{eq:convol-moment}) applies to any hard-scattering function, whether derived by dQCD or effective theory.
Then, if different resummations for $C(z)$ agree at the level of moments (which we have seen to be the case when we choose $\mu_s=M/\bar s_1$ above), they should
agree phenomenologically.\footnote{ A technical observation is that the dQCD ``minimal prescription" described above in connection with Eq.\ (\ref{eq:qcd-min0}), which is taken to the left of the Landau pole in $N$ space,  does not strictly speaking result in a convolution in the form of Eq.\ \eqref{eq:convol-moment}  We can, however, safely neglect such exponentially-suppressed corrections in this discussion, and we will confirm that they are negligible in the cases we consider.}   

To estimate corrections, we consider the Taylor expansion of $\ln {\cal L}(\tau/z)$, the logarithm of the luminosity, around partonic threshold, $z=1$,
 \bea
\ln  {\cal L}\left(\frac \tau z\right)  = \sum_{n =0}^\infty \frac{1}{n!} s_n(\tau)\, \ln^n z\, ,
 \label{eq:L-expand}
 \eea
 with 
 \bea
 s_n(\tau) =  (-1)^n\, \frac{d^n \ln{\cal L}(\tau')}{d\ln^n\tau' } \Bigg |_{\tau'=\tau}\, .
  \label{eq:sn-def}
 \eea
 Now  $s_1$ describes the power behavior of ${\cal L}$ when $z$ is close enough to $z=1$.  This single-power approximation to the luminosity will give a good approximation to the entire resummed cross section if ${\cal L}(\tau/z)$ becomes small by the time the higher order terms proportional to $s_n$, $n\ge 2$ become comparable
 to the first term in the expansion.
 
 To get a more quantitative sense of the requirements for the single power approximation, we note that the luminosity ${\cal L}$ decreases by a factor $(1/e)^p$ when $\ln \frac{1}{z} \sim p/s_1(\tau)$.    For the power $z^{s_1}$ to dominate until ${\cal L}$ becomes small, we need 
 \bea
 \left | r_n\right | \ll 1\, , \quad
 r_n \equiv \frac{1}{n!} \frac{s_n(\tau)}{\left[s_1(\tau)\right]^n}\, ,  \quad n\ge 2\, ,
 \label{eq:s_n-inequal}
 \eea
 for all relevant values of $n$.
  We will argue that this is a common feature of realistic parton distributions.
 
A particularly simple example is a model function that reflects the steep
decline of parton luminosities \cite{Appell:1988ie,Catani:1998tm,Becher:2007ty}, as
\begin{equation}
{\cal L}(\tau') = A (1-\tau')^\beta,
\label{eq:L-simple}
\end{equation}
for $\beta \gg 1$. From Eq. \eqref{eq:sn-def} we have for the first logarithmic derivative
\bea
s_1(\tau) =\beta\, \tau / (1-\tau)\, .
\eea
 Clearly, the $s_n$ remain linear in $\beta$ when $n \geq 2$. Therefore $r_n$, the measure of convergence in Eq. \eqref{eq:s_n-inequal}, is proportional to $\beta^{-(n-1)}$, which is generically suppressed.
 Indeed, as long as $\beta \tau>1$, the logarithmic expansion \eqref{eq:L-expand} of the luminosity converges, and we expect $|r_n|\ll 1$ to hold quite generally, for $\tau>1/\beta$, not only for $\tau\rightarrow 1$.   Assuming this to be the case, Eq.\ (\ref{eq:convol-moment}) is a good approximation, and $d\sigma/d M^2$ is proportional to a specific moment of the hard-scattering cross section.  In the effective theory treatment, a scale choice that eliminates large logs from the soft function is $\mu_s = M/\bar s_1 = M(1-\tau)/(\beta\,\tau e^{\gamma_E})$,  similar to the estimate of Ref.\ \cite{Becher:2007ty} for this same luminosity.
 
 More realistic models of parton luminosities can be constructed from model parton distributions like
 \bea
 f(x) = C\, x^{-\delta} \, \left(1-x\right)^\beta\, .
 \label{eq:f-power}
 \eea
 We can test the single-power approximation explicitly in this case as well, using that
the convolution of two such distributions gives a luminosity of hypergeometric form,
 \bea
 {\cal L}(\tau') = C^2\, B(\beta+1,\beta+1) \, \tau'{}^{-\delta}\, \left(1-\tau'\right)^{2\beta+1}\, F_{2,1}\left(\beta+1,\beta+1;2\beta+2;1-\tau'\right)\, .
 \label{eq:L-example}
 \eea
We show in Appendix \ref{sec:appendix} that this luminosity  is again dominated by a single power, both for small $\tau$, even with a moderate power, $\delta = {\cal O}(1)$, and
 for all $\tau$ when $\beta\gg 1$.  
 
\begin{figure}
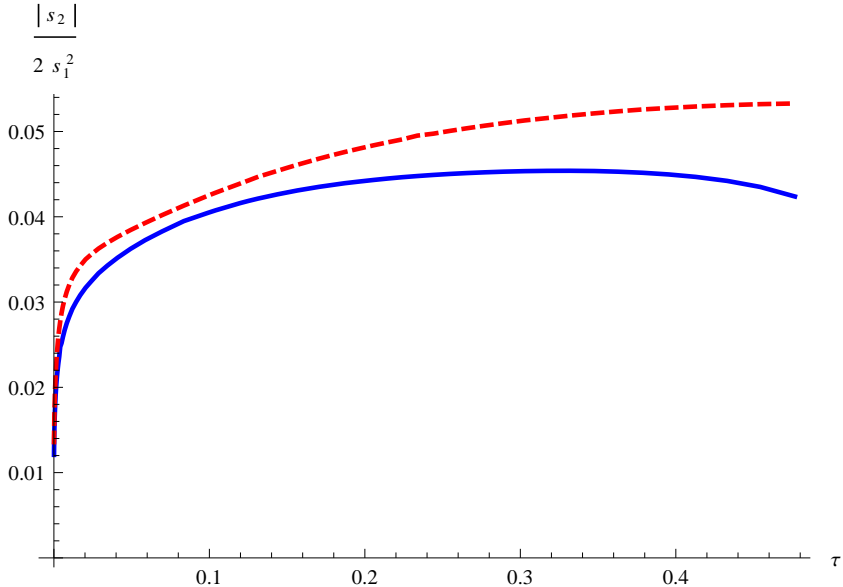

\centerline{ \figscale{ratio}{11cm}}
\caption{The ratio $\left|r_2\right|$, Eq.\ (\ref{eq:s_n-inequal}) , as a function of $\tau$ for the gluon distributions of Fig.\ \ref{fig:tau-singlepower}.
\label{fig:ratio}}
\end{figure}

\begin{figure}
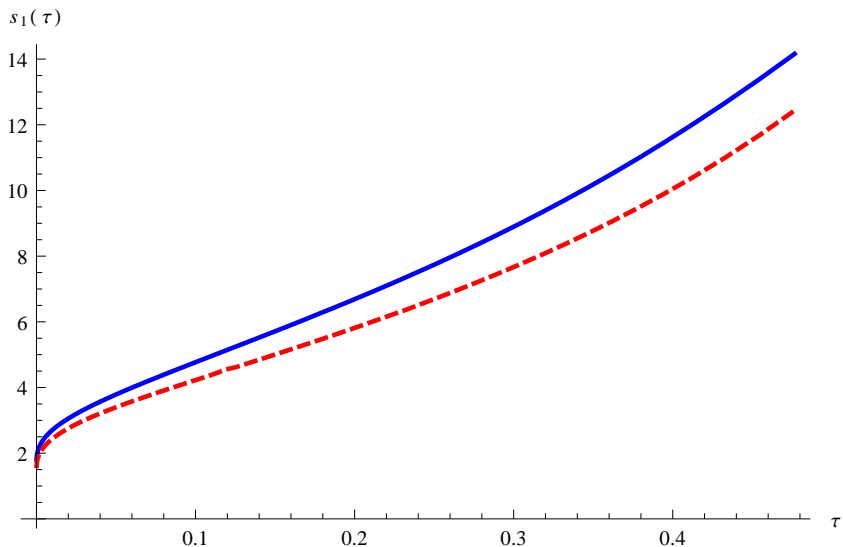

\centerline{ \figscale{s-effective}{11cm}}
\caption{  The first logarithmic derivative, $s_1$, Eq.\ (\ref{eq:sn-def}), as a function of $\tau$ for the gluon distribution of Fig.\ \ref{fig:tau-singlepower}.  \label{fig:s-eff}}
\end{figure}

 We next test condition (\ref{eq:s_n-inequal}) for $n=2$, $|r_2|\ll 1$, for a fully realistic parton distribution.
   We show in Fig.\ \ref{fig:ratio} $|r_2|=|s_2|/(2 s_1^2)$ for the gluon-gluon luminosity as a  function of $\tau$ for $\mu_f=10$ GeV and $126$ GeV.   In both cases the ratio satisfies $\left|r_2 \right| \ll 1$ for all $\tau$ up to 0.5, even where $s_1$ is not large.  The slope itself, $s_1$, is shown in Fig.\ \ref{fig:s-eff}. Luminosity functions arising from valence quark distributions, for example, the $u \bar u$ luminosity function, also satisfy $|r_2|\ll 1$. It is therefore natural to conclude that for many, probably most, cases of phenomenological interest, the effective power $s_1(\tau)$ determines the dominant moment, and  consequently an effective soft scale, $\mu_s= M/\bar s_1(\tau)$.

\begin{figure}
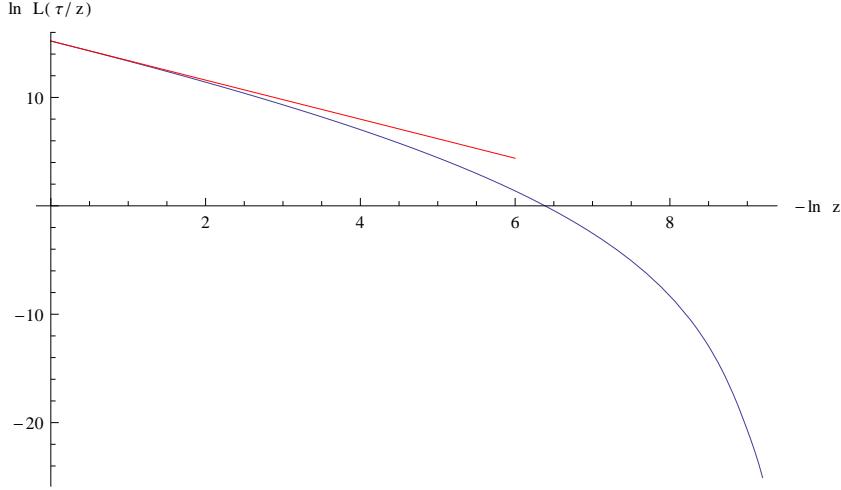

\centerline{ \figscale{gg_Higgs_lum}{11cm}}
\caption{Logarithmic plot of gluon luminosity with its single-power approximation approximation \label{fig:log-plot}}
\end{figure}

In Fig.\ \ref{fig:log-plot}, which is analogous to Fig.\ 1 of \cite{Becher:2007ty},
we give a logarithmic plot of the gluon-gluon luminosity versus $\ln(\tau_0/z)$, with $\tau_0 = (126 \ {\rm GeV}/ 8 \ {\rm TeV})^2 \approx (1/8)^4$, characteristic of 126 GeV (=$\mu_f$) Higgs boson production in proton-proton collisions at center-of-mass energy 8 TeV.  The linear approximation is accurate while the luminosity decreases over several factors of $e$.
In this case, the first derivative in the Taylor expansion \eqref{eq:L-expand} evidently dominates the second over an adequate region, even though the first derivative itself is not very large.

\subsection{Corrections to the single power approximation}

Having shown the equality of dQCD and SCET threshold resummations in the single-power approximation, we turn to corrections.   These are associated with coefficients, $s_n$, $n>1$ in Eq. \eqref{eq:L-expand}, in terms of which the luminosity function can be written as
\begin{align}
{\cal L} \left( \frac \tau z \right) &= {\cal L} \left( \tau \right) z^{s_1} \,\exp \left[\frac 1 2 s_2 \ln^2 z + \frac 1 6 s_3 \ln^3 z + \frac 1 {24} s_4 \ln^4 z + \dots \right] \nonumber \\
&= {\cal L} \left( \tau \right) z^{s_1} \, \left[1 + \frac 1 2 s_2 \ln^2 z + \frac 1 6 s_3 \ln^3 z + \left( \frac 1 8 s_2^2 + \frac 1 {24} s_4 \right) \ln^4 z + \dots \right].
\end{align}
This expansion enables us to refine the expression for the hadronic cross section in Eq. \eqref{eq:convol-moment} as
\begin{align}
\frac{d \sigma (\tau)}{d M^2} &= \sigma_0 \int_{\tau}^1 \frac{dz}{z} {\cal L} \left( \frac \tau z \right) C(z) \nonumber \\
&= \sigma_0 \int_0^1 \frac{dz}{z} {\cal L} \left( \frac \tau z \right) C(z)\  +\  {\cal O}\left({\cal L}(\tau)\tau^{s_1}\right) \nonumber \\
&= \sigma_0 \,{\cal L} \left(\tau\right) \int_0^1 \frac {dz} {z} z^{s_1} \, \left[1 + \frac 1 2 s_2 \ln^2 z 
+ \cdots \right] C(z) \nonumber \\
&= \sigma_0 \,{\cal L} \left(\tau\right) \left[ \widetilde C \left(s_1\right) + \frac 1 2 s_2 \,\widetilde C'' \left(s_1\right) 
+ \cdots \right]\, ,
\label{eq:convol-expansion}
\end{align}
where for simplicity we keep only $s_1$ and $s_2$ terms in the expansion. Using
\begin{equation}
\left( \frac{d}{d s_1} \right)^2 = \frac {1} {s_1^2} \left[ \left( \frac{d}{d \ln s_1} \right)^2 - \frac{d}{d \ln s_1} \right],
\end{equation}
we have
\begin{align}
\frac{d \sigma (\tau)}{d M^2}  = \sigma_0 {\cal L} \left(\tau\right) \left \{ 1 + r_2\left[ \left( \frac{d}{d \ln s_1} \right)^2 - \frac{d}{d \ln s_1} \right] + \dots \right \} \widetilde C \left(s_1\right)\, ,
\label{eq:convol-expansion-2}
\end{align}
where $r_2={s_2}/{2 s_1^2}$ as in Eq.\ (\ref{eq:s_n-inequal}).
Corrections to the single-moment approximation in Eq.\ (\ref{eq:convol-moment}) are thus suppressed by powers of $\ln s_1$ as well as by the coefficients $r_n$, defined in (\ref{eq:s_n-inequal}).
The convergence of Eq.\ \eqref{eq:convol-expansion} as a  series is thus quantifiable.  

We can also test differences between SCET and dQCD in the new expansion.
Applying \eqref{eq:convol-expansion} to the difference between SCET and dQCD resummed moments with soft scale $\mu_s = M / \bar s_1$ in the former, we obtain
\begin{align}
&\hspace{-10mm} \frac{1}{\sigma_0 {\cal L}(\tau)} \left[ \frac{d \sigma (\tau)}{d M^2}{}^\sc \left( \tau, \mu_s=\frac{M}{\bar s_1} \right) - \frac{d \sigma (\tau)}{d M^2}{}^\dq (\tau) \right]
\nonumber \\
& \approx \left \{ 1 + r_2 \left[ \left( \frac{d}{d \ln s_1} \right)^2 - \frac{d}{d \ln s_1} \right] \right \} \left[  \widetilde{C}^{\rm [SCET]} \left( s_1, \mu_s=\frac{M}{\bar s_1} \right) -  \widetilde C^{\rm [dQCD]} (s_1) \right] \nonumber \\
&= \left \{ 1 + r_2 \left[ \left( \frac{d}{d \ln s_1} \right)^2 - \frac{d}{d \ln s_1} \right] \right \} \left[ \widetilde{C}_r^{-1} \left(\ln \frac {M}{\bar s_1 \mu_s}, \as(\mu_s) \right) - 1 \right] \widetilde C^{\rm [dQCD]} \left(s_1\right),
\label{eq:difference-corrections}
\end{align}
where we have used the moment ratio Eq.\ \eqref{eq:Cr} in the second equality. Recall from Sec.\ \ref{sec:partonic} that the moment ratio $\widetilde C_r$ deviates from unity only starting at  order $\as^k$ in the case of ${\rm N}^k$LL resummation, when the soft function is computed to order $\as^{k-1}$.   We can then expand (\ref{eq:difference-corrections}) as
\begin{align}
&\quad\, \frac{1}{\sigma_0 {\cal L}(\tau)} \left[ \frac{d \sigma (\tau)}{d M^2}^\sc \left( \tau, \mu_s=\frac{M}{\bar s_1} \right) - \frac{d \sigma (\tau)}{d M^2}^\dq (\tau) \right] \nonumber \\
& \hspace{10mm} \approx \left \{ 1 + r_2 \left[ \left( \frac{d}{d \ln s_1} \right)^2 - \frac{d}{d \ln s_1} \right] \right \} \sum\limits_{m \geq k} \,\sum\limits_{1 \leq n \leq 2m} A_{mn} \as^m \ln^n \frac{M}{\bar s_1 \mu_s} \widetilde C^\dq \left(s_1\right)\, |_{\mu_s=M/s_1}
\nonumber \\
&\hspace{10mm} = \frac{s_2}{2 s_1^2} \sum\limits_{m \geq k} \,\sum\limits_{1 \leq n \leq 2m} \left( 2A_{m2} + A_{m1}\right)\, \as^k\, \widetilde C^\dq\left(s_1\right)\, .
\label{eq:s2-difference}
\end{align}
In the final form, only the $n=1,\,2$ terms give non-zero contributions when $\mu_s = M/\bar s_1$.  The leading term in Eq. \eqref{eq:s2-difference} is thus
of order ${\rm N}^{k+1} {\rm LL}$, that is, subleading compared to the original calculation, in addition to its suppression by $r_2$.
This argument generalizes straightforwardly  to  coefficients $s_n$ at any order.  It implies that the difference between dQCD and SCET resummed hadronic cross sections is of subleading logarithmic order, and further suppressed in the luminosity expansion.   This will be the case whenever the perturbation expansion Eq. \eqref{eq:convol-expansion} is justified by the condition Eq. \eqref{eq:s_n-inequal}, which we have verified  for a variety of model and realistic parton distributions.

\subsection{Cross section comparisons}

From the discussion above, we expect that pQCD and SCET resummations should give very similar numerical results, at least when the soft scale in the latter is chosen as $M/\bar s_1$.  To test this, we compare dQCD and SCET resummations directly in a specific case, designed to give a controlled numerical comparison between the two.  To eliminate differences at non-leading powers of $1-z$, we invert the Mellin transform numerically for both  dQCD and SCET as in Eq.\ (\ref{eq:mellin-inversion}), after multiplying by the moment of a fitted parton luminosity function \cite{deFlorian:2007sr}.  For the dQCD form, we use the contour specified by the minimal prescription.  

We use Eq.\ \eqref{eq:soln-scet-2} as the definition of the SCET coefficient function in moment space.\footnote{The various functions, such as $S_{\rm cusp}$ and $\eta$ are found in the Appendix B of \cite{Becher:2007ty}.}  We then use the identity $\tilde C^\dq \left(\ln \frac {1}{\bar N}, \as(M)\right)= \tilde C^\sc \left(\ln \frac {1}{\bar N}, \as(M), \mu_s=\frac{M}{\bar N} \right)$,  Eq.\ (\ref{eq:equality}), as the defintion of the dQCD resummed moment, again using Eq.\ \eqref{eq:soln-scet-2} to ensure that this relation is exactly satisfied  in our numerical implementation, free of differences at subleading powers of $1/N$ or $1-z$ and subleading logarithms.\footnote{An example of the latter are differences between the two expressions on the right-hand side of Eq.\ (\ref{eq:S-soln-qcd-1}) when $\gamma_W$ and $\hat D$ are truncated to finite order.}  Again, for the SCET soft scale we use $\mu_s = M/\left(s_1 e^{\gamma_E}\right)$, with $s_1$ determined numerically from the luminosity.  Otherwise, the moment inversion here is completely independent of the single power approximation.   We emphasize that with this (or indeed any) $N$-independent choice of $\mu_s$, the SCET and dQCD coefficient functions are still completely different for orders beyond $\as^2$ in the expansion of their corresponding NNLL resummations.

The resummations are performed to NNLL accuracy in the dQCD convention, that is, taking $s_{\rm DY}$, $\hat D$ and $\gamma_W$ in Eq.\ (\ref{eq:soln-scet-2}) to order $\as^2$, $\Gamma_{\rm cusp}$ to order $\as^3$, and using the $3$-loop running coupling.
The following results are evaluated with only the resummed soft function.  We do not include constant factors such as the hard function with the effective top vertex matching coefficient, and do not perform matching to fixed order. Therefore, the plots we show are not intended as phenomenological predictions, but to serve as illustrations for the points we have made previously.

\begin{figure}
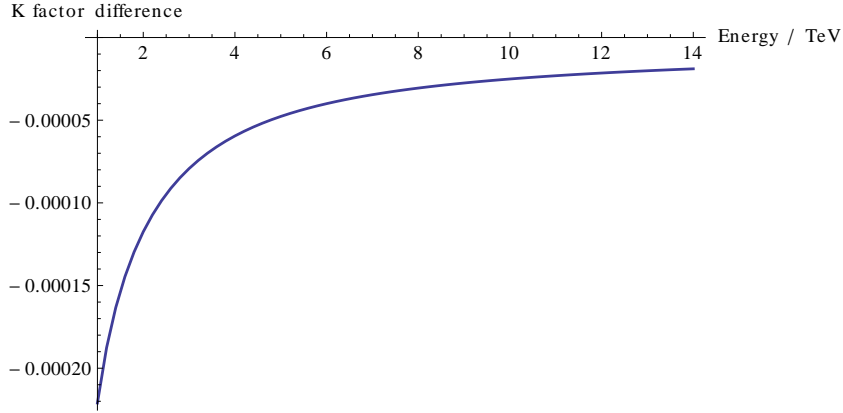

\centerline{\figscale{SCET_dQCD_diff}{11cm} }
    \caption{Relative difference between SCET resummation with $\mu_s = M / \bar s_1$ and dQCD resummation for the 126 GeV Higgs K factor, at hadronic collisin energies from 1 to 14 TeV.}
	\label{fig:pure-resum3}
\end{figure}

Fig. \ref{fig:pure-resum3} shows the fractional difference between SCET and dQCD resummed K-factors (ratios to the Born process) for the same gluon distributions as above. We can see that the difference is tiny, below $10^{-4}$ for LHC energies.  This can be understood because first, the ratio $r_2=s_2/(2 s_1^2)$ that characterizes corrections to the single-power approximation  is quite small for small $\tau$, as illustrated in Figs. \ref{fig:ratio} and \ref{fig:log-plot}, and second, its coefficient is of subleading logarithmic order (N$^3$LL in this case).  This is also consistent with the observation in Ref.\ \cite{Ahrens:2008nc}, where much of the numerical difference between dQCD and SCET resummed Higgs cross sections was traced to non-leading powers of $1-z$ arising from the different Mellin inversion methods.    Computing both dQCD and SCET cross sections numerically eliminates such differences, and the underlying difference between the two methods turns out to be very small in this case.   

The consistency of dQCD and SCET results again suggests that the single power approximation should work well for either.   Indeed, this turns out to be the case.  Fig.\ \ref{fig:pure-resum1}a shows K-factors for the gluon fusion production of a 126 GeV Higgs, at 1-14 TeV energies. The upper, dashed  curve shows the K factor from the single power approximation, that is, $\tilde C_{\rm dQCD} \left(N=s_1(\tau),M\right)$. The lower, solid  curve shows the dQCD prediction using the full parton distribution. The full dQCD resummation agrees with the single power approximation very well, even before the corrections of Eq.\ (\ref{eq:convol-expansion-2}) to the latter.   When the  correction associated with $s_2$ in Eq.\ (\ref{eq:convol-expansion-2}) is taken into account, the agreement is impressive, as shown by the difference between the corrected single power and full dQCD predictions plotted in Fig.\ \ref{fig:pure-resum1}b.

\begin{figure}
\centerline{\figscale{Higgs_single_power}{12cm} }
\centerline{ (a) }

\vbox{\vskip .25 in}

\centerline{ \figscale{Single_power_error}{12cm}}
\centerline{(b) }
    \caption{(a) Plot of 126 GeV Higgs resummed gluon fusion K factor, omitting the hard function, from dQCD and the corresponding single power approximation without $s_2$ correction.   (b) Difference between same full resummed K factor and corresponding single power correction with its $s_2$ correction.}
	\label{fig:pure-resum1}
\end{figure}

In summary,  for Higgs production we have verified both the usefulness of the single-power approximation, and the consequent consistency of the dQCD and SCET resummation formalisms.  Of course, the full implementation of these formalisms requires the complete hard function and matching to fixed-order results, but this will not affect the conclusion that the intrinsic difference between the two resummation methods is tiny in the case of 126 GeV Higgs. Although we do not attempt a phenomenological study of other cases here, the analytic arguments and parton luminosity plots in the previous subsections seem strongly to suggest consistency between the two resummation methods over a wide kinematic range.  They suggest as well that the single power approximation often may be a useful tool in its own right.

\subsection{Soft scale comparisons}
A widely used procedure in the SCET threshold resummation literature is minimizing the correction from the one-loop soft function after convoluting with the parton luminosity function \cite{Becher:2007ty,Ahrens:2010zv}. The method of this paper suggests an alternative choice, $\mu_s = M / \left[ s_1(\tau) e^{\gamma_E} \right]$, based directly on the shape of the parton luminosity function. A previous proposal, also based on the parton luminosity shape, in Ref. \cite{Bauer:2010jv} suggests $\mu_s = \lambda^2 M$, where $\lambda^2$ is defined by the consistency equation
\begin{equation}
\lambda^2 = \frac {\int_\tau^{1-\lambda^2} \frac{dz}{z} \mathcal L \left( \frac \tau z \right)} {\int_{1-\lambda^2}^1 \frac{dz}{z} \mathcal L \left( \frac \tau z \right)}.
\end{equation}
\begin{figure}
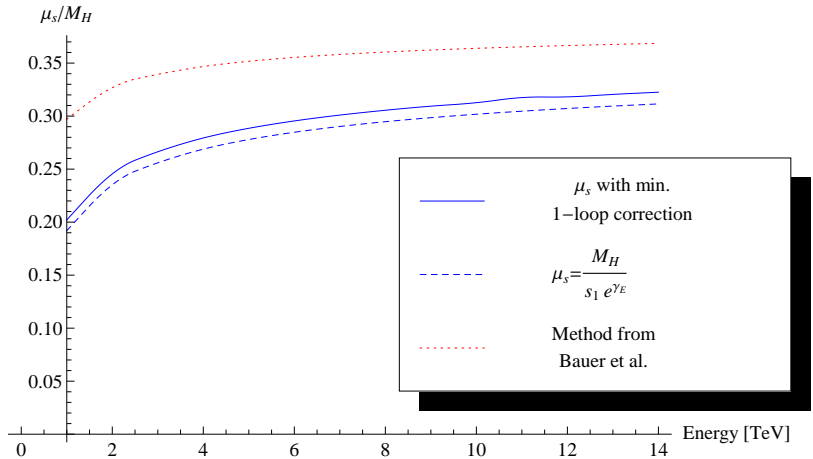

\centerline{\figscale{softscale}{11cm} }
    \caption{Soft scales for inclusive Higgs production obtained from minimizing the one-loop soft correction \cite{Ahrens:2008nc} (using contour Mellin inversion), from the $s_1(\tau)$ variable in the method of this paper, and from a method proposed in \cite{Bauer:2010jv}.}
	\label{fig:soft-compare}
\end{figure}
We plot in Fig. \ref{fig:soft-compare} the soft scales $\mu_s$ obtained from the three methods, for a 126 GeV Higgs produced from gluon fusion at hadronic collision energies from 1 TeV to 14 TeV. We see that the method of this paper agrees with the method of minimizing one-loop soft corrections within a few percent, and also agrees well with the method from \cite{Bauer:2010jv}, given that, in practice, scales are varied up and down by a factor of 2, which is much larger than the differences shown in the figure.

For Higgs and Drell-Yan, the one-loop soft function is proportional to $2 \ln^2 [M^2 / (\bar N^2 \mu_s^2)] + \pi^2 / 3$ \cite{Becher:2007ty, Ahrens:2008nc}, which is clearly minimized at $\mu_s = M / \bar s_1$ to the extent that the single-power approximation is valid. This explains the close agreement we observe.

\subsection{Threshold suppressed differences}
As shown in \citep{Ahrens:2008nc}, the choice of $(1-z)$-suppressed terms in the widely used inversion formula in SCET differs from exact Mellin invesion, to a first approximation, by an extra factor of $1/\sqrt z$, accounting for more than 70\% of the difference between threshold enhancements for the Higgs total cross section from dQCD and SCET threshold resummations. In the framework of the single power approximation, this replaces the power $s_1$ in Eq. \eqref{eq:convol-moment-2} by $s_1 - 1/2$, giving us a result that is approximated by $\tilde C(s_1 - 1/2)$ instead of $\tilde C(s_1)$.
\begin{figure}
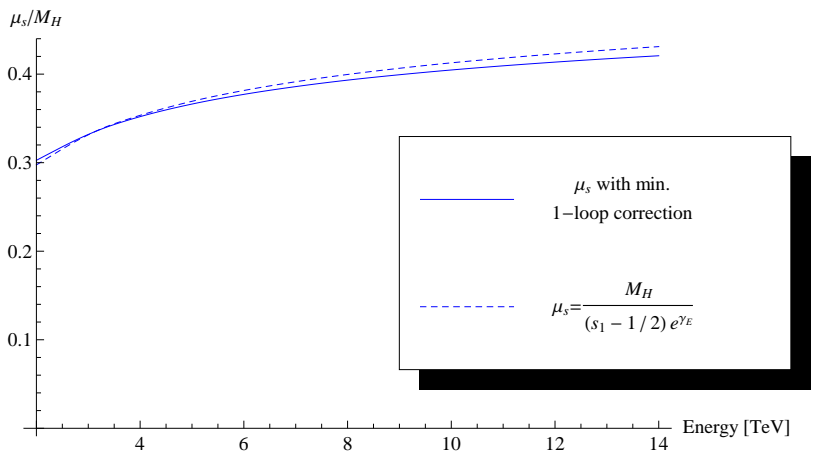

\centerline{\figscale{softscaleZ}{11cm} }
    \caption{Soft scales obtained from minimizing the one-loop soft correction using the SCET inversion formula of \cite{Becher:2007ty}, and from the $s_1(\tau) - 1/2$ variable in the method of this paper, for the production of a 126 GeV Higgs at various energies.}
	\label{fig:soft-compare-shifted}
\end{figure}
In Fig. \ref{fig:soft-compare-shifted} we plot the comparison between the modified soft scale we predict, $\mu_s = M / \left[ (s_1 - 1/2) e^{\gamma_E} \right]$, and the soft scale found by minimizing the one-loop soft correction using the widely adopted SCET inversion formula instead of contour Mellin inversion. The modified single power approximation agrees with the result of minimizing the one loop correction \cite{Ahrens:2008nc} within a few percent.

Given that $s_1$ is often not very large (about 1.8 for a 126 GeV Higgs from 14 TeV collisions), this downshift of $s_1$ by $1/2$ should generally lead to milder threshold enhancement. This effect is present in Ref. \cite{Ahrens:2008nc} for Higgs production, and in Ref. \cite{Becher:2007ty} for the Drell-Yan process in certain parameter regions. An important dependence on threshold-suppressed factors has also been observed in the recent calculation of the gluon fusion Higgs cross-section at ${\rm N}^3$LO \cite{Anastasiou:2014vaa} in the threshold limit.

\section{Conclusion}

We have explored the similarities of dQCD and SCET threshold resummations, in the implementations of Refs.\ \cite{Catani:2003zt} and \cite{Becher:2007ty}, respectively, and reaffirmed that they share common properties of factorization and evolution, differing primarily in the use of boundary conditions for the evolution.  For the dQCD formalism, evolution starts in moment space at scale $M/\bar N$, and in the SCET formalism at a soft scale, $\mu_s$, determined by the interplay of parton luminosities and the hard scattering function. We  presented a new analysis of the dQCD resummed coefficient expressed as the exponential of a Mellin transform, Eq.\ \eqref{eq:qcdmoment}, showing that the non-leading resummation function $D(\as)$ is a specific sum of web diagrams computed in momentum space.  We went on to extend the moment ratio method of Ref.\ \cite{Bonvini:2012az}, confirming that the two resummation formalisms give partonic cross sections that agree to any desired logarithmic order when the SCET boundary condition for evolution is chosen as $\mu_s=(1-z)M$.  
  
  For hadronic cross sections, we found that for a wide range  of $\tau$, the cross section is dominated by a single Mellin moment, determined by logarithmic derivatives of the luminosity function.   Corrections to this approximation can be computed by an expansion of the logarithm of the luminosity around its value at threshold.  We expect that for most collider scenarios, dQCD resummations, using minimal or other prescriptions, should be close to SCET threshold resummation when the soft scale of the latter is close to the value determined as above.    The single-power approximation, including corrections, can provide a useful simplification in many cases,  without sacrificing numerical accuracy. The approximation also simplifies the determination of the soft scale and the analysis of the effect of threshold-suppressed terms.

\acknowledgments
We thank Andrea Banfi, Sally Dawson, Andrea Ferroglia, Stefano Forte, Chris Lee, Ben Pecjak, Frank Petriello, Frank Tackmann and Jonathan Walsh for discussions.   
In particular, we thank Leandro Almeida, Steve Ellis, Chris Lee, Ilmo Sung and Jonathan Walsh for extensive discussions on  the relationship between dQCD and SCET treatments of infrared safe jet shapes.
This work was supported in part by the National Science Foundation,  grants  PHY-0969739 and -1316617.

\appendix
\section{Relating different forms of the non-leading resummation function}
\label{sec:appendix}

At first sight, the moment space form of the resummed exponent, Eq.\ \eqref{eq:qcdmoment} looks rather different than the exponent in Eq.\ (\ref{eq:soln-gen}),  found by solving the soft function evolution equation.   
To leading power in $N$, however, the Mellin expression in Eq. \eqref{eq:qcdmoment} can be rewritten as a double integral in which $N$ appears only in the limits of integration, acted upon by an infinite series of derivatives, 
\begin{align}
\label{eq:appendix-qcd-mom}
E \left(N,M\right) &= -\int_0^1 dz \frac{z^{N-1}-1} {1-z} \left[ 4\int_{(1-z) M}^{M} \frac{d\mu'}{\mu'} A \left(\as\left(\mu'\right)\right) - D \left( \as\left( (1-z) M \right) \right) \right] 
\nonumber \\
& \hspace{-10mm}= \sum_{k=0}^\infty
\frac{\Gamma^{(k)}(1)}{k!}
\frac{d^k}{d\ln^k \frac 1 N}\int_0^{1-1/N} \frac{dz}{1-z}\, \left[ 4\int_{(1-z) M}^{M} \frac{d\mu'}{\mu'} A \left(\as\left(\mu'\right)\right) - D \left( \as\left( (1-z) M \right) \right) \right] 
\nonumber\\
&\ \hspace{20mm} 
+\ {\cal O}\left(\frac{1}{N}\right)
\nonumber\\
& \hspace{-10mm} \equiv \Gamma\left( 1+ 2\frac{d}{d\ln\frac{1}{N^2}}\right)\, \int_0^{1-1/N} \frac{dz}{1-z}\, \left[ 4\int_{(1-z) M}^{M} \frac{d\mu'}{\mu'} A \left(\as\left(\mu\right)\right) - D \left( \as\left( (1-z) M \right) \right) \right] 
\, ,
\end{align}
where we have set $\mu_f=M$ for simplicity. 
In the second equality we have replaced the explicit Mellin moment by a modified upper limit in the $z$ integral, using the identity,
\begin{equation}
\int_0^1 dz\,z^{N-1}\,\left[\frac{\ln^p(1-z)}{1-z}\right]_+
=
-\sum_{k=0}^{p+1}
\frac{\Gamma^{(k)}(1)}{k!}
\frac{d^k}{d\ln^k \frac 1 N}\int_0^{1-1/N}dz\,\frac{\ln^p(1-z)}{1-z}
+\mathcal{O} \left(\frac{1}{N}\right)\, .
\label{eq:MellinGamma}
\end{equation}
This relation holds for any power of $\ln(1-z)$, up to inverse powers of $N$ as indicated.  It therefore applies to the full $\mu'$ integral in (\ref{eq:appendix-qcd-mom}), which is an expansion in  logarithms of $1-z$ only.   A proof of (\ref{eq:MellinGamma}) follows from using $(1-z)^\delta$ as a generating function for powers of $\ln(1-z)$, as in Ref.\ \cite{Forte:2002ni}.   Following Refs.\ \cite{Catani:2003zt} and \cite{Becher:2007ty}, we now go on to review how the action of the derivatives can be absorbed into a modified function $D(\as)$, as in Eq.\ (\ref{eq:relationD}).

To make closer contact with the alternate form of the exponent, we change variables in Eq.\ (\ref{eq:appendix-qcd-mom}) to $\mu = (1-z)M$, giving
\bea
E(N,M) = \Gamma\left( 1+ 2\frac{\partial }{\partial \ln\left(\frac{M}{N}\right)^2}\right)\, \int^M_{M/N}\frac{d\mu}{\mu} \, \left[ 4\int_\mu^{M} \frac{d \mu'}{\mu'} A \left(\as\left(\mu'\right)\right) - D \left( \as\left( \mu \right) \right) \right] 
\, ,
\label{eq:E-NM}
\eea
again, up to power corrections in $N$.   Here, we have converted the derivative with respect to $\ln N$ to a derivative with respect to $\ln (M/N)$, treating $M$ and $M/N$ as independent.
To streamline subsequent expressions, we define a function that represents the integrals on which the derivatives in Eq.\ (\ref{eq:MellinGamma}) act,
\begin{equation}
\label{eq:appendix-qcd-I}
 I \left(\rho,M \right) = \int_{\rho}^{M} \frac {d\mu}{\mu} \left[ \int_{\mu}^{M} \frac {d\mu'}{\mu'} 4A\left(\as\left(\mu'\right)\right) - D \left(\as\left(\mu\right)\right) \right]\, ,
\end{equation}
where in (\ref{eq:E-NM}), $\rho=M/N$.    We now introduce the notation \cite{Becher:2007ty},
\bea
\nabla  \equiv \frac{d}{d \ln \rho^2} = \frac{\beta\left(\as\left(\mu'\right)\right)}{2}\, \frac{d}{d\as(\mu')}\, ,
\label{eq:nabla-def}
\eea
 and use $2e^{\gamma_E \nabla}$ to translate the lower limit of the $\mu$ integral in Eq.\ (\ref{eq:appendix-qcd-I}),
\begin{align}
\label{eq:appendix-qcd-mom1}
E \left(\frac{M}{N},M\right) &= \Gamma \left(1+2\nabla\right)  I \left(\frac{M}{N},M\right) \nonumber \\
&= \left[e^{\gamma_E 2\nabla} \Gamma \left(1+2\nabla\right) \right] \, e^{-\gamma_E 2\nabla} \, I \left(\frac{M}{N},M\right)
\nonumber\\
&= \left[e^{\gamma_E 2\nabla} \Gamma \left(1+2\nabla\right) \right] I \left(\frac{M}{\bar N},M\right)\, ,
\end{align}
where, as above, $\bar N \equiv e^{\gamma_E} N$.   The expansion
\bea
e^{2\gamma_E \nabla} \Gamma \left(1+2\nabla\right) &=& 1 + \frac{\pi^2}{12} (2\nabla)^2 - \frac 1 3 \zeta\left(3\right) (2\nabla)^3 \ +\ \cdots
\nonumber\\
&\equiv& 1 + \sum_{k=2}^\infty a_k (2\nabla)^k
\eea
 has no linear term in $\nabla$.
Carrying out the derivatives of Eq. \eqref{eq:E-NM} in this notation, we find 
\bea
e^{2\gamma_E \nabla} \Gamma \left(1+2\nabla\right) I \left(\frac{M}{\bar N},M\right)  &=&
I \left(\frac{M}{\bar N},M\right) 
\nonumber\\
&\ & \hspace{-25mm}+ \sum_{k=2}^\infty a_k \left[ 4\nabla^{k-2} A\left(\as\left( {M}/{\bar N} \right)\right) + \nabla^{k-1} D\left(\as\left( {M}/{\bar N} \right)\right) \right]\, ,
\label{eq:derivative-terms}
\eea
which  depends on $M$ only through the running coupling $\as(M/N)$.    The derivatives $\nabla$ may therefore also be converted to
derivatives with respect to the coupling, as in (\ref{eq:nabla-def}) above.
With these results in hand, we   rewrite the derivative terms in (\ref{eq:derivative-terms}) in a way that reintroduces an integral over the scale of the running coupling,
\begin{align}
\label{eq:appendix-derivative}
4\nabla^{k-2} A\left(\as\left( {M}/{\bar N} \right)\right) + \nabla^{k-1} D\left(\as\left( {M}/{\bar N} \right)\right) &= 4\nabla^{k-2} A\left(\as\left( M \right)\right) +\nabla^{k-1} D\left(\as\left( M \right)\right) \nonumber \\
& \hspace{-30mm} + \int_{M/\bar N}^{M} \frac {d\mu'}{\mu'} \left[-4\nabla^{k-1} A\left(\as\left( \mu' \right)\right) -\nabla^{k} D\left(\as\left( \mu' \right)\right) \right]\, .
\end{align}
In this expression, of course, the $M$ dependence of the constant term is canceled by the upper limit of the integral.   Thus, up to an $\bar N$-independent constant that only affects the overall cross section, substituting Eq.\ \eqref{eq:appendix-derivative} into the final expression for the exponent $E$ in Eq.\ \eqref{eq:appendix-qcd-mom1} gives
\begin{align}
\label{eq:qcdcutoff}
E \left(\frac{M}{N},M\right)  &\cong \int_{M/\bar N}^{M} \frac {d\mu}{\mu} \left[ \int_{\mu}^{M} \frac {d\mu'}{\mu'} 4A\left(\as\left(\mu'^2\right)\right) - \hat{D} \left(\as\left(\mu^2\right)\right) \right] \nonumber \\
&= \int_{M/\bar N}^{M} \frac {d\mu}{\mu} \left[ -4A\left(\as\left(\mu\right)\right) \left(\ln \frac{M}{N\mu}  \right) - \hat{D} \left(\as\left(\mu\right)\right) \right]\, ,
\end{align}
where we drop the ``constant" terms that depend only on $\as(M)$, and where in the second expression we first perform the $\mu$ integral, and then relabel $\mu'$ as $\mu$.    A modified $D$ term, $\hat D$, has been introduced to absorb the effects of all derivatives $\nabla^k$, $k\ge 2$ in the expansion of Eq.\ (\ref{eq:appendix-qcd-mom1}), where
\begin{align}
 \hat{D}\left( \as\left(\mu\right) \right) &\equiv e^{2\gamma_E \nabla} \Gamma \left( 1+2\nabla \right) D\left( \as\left(\mu\right) \right) + \frac {e^{2\gamma_E \nabla} \Gamma \left( 1+2\nabla \right) - 1}{\nabla} A \left( \as\left(\mu\right) \right)\, ,
\end{align}
the result given in Ref.\ \cite{Catani:2003zt} .

\section{The single power approximation with model parton distributions}

In this appendix, we confirm explicitly the single-power expansion for parton distributions of the general form, $x^{-\delta}(1-x)^\beta$, $\delta,\ \beta > 0$.
As mentioned in Section \ref{subsec:single}, the convolution of two such parton distribution functions, gives a parton luminosity function of the form
\begin{equation}
{\cal L}(\tau) = C^2 \tau^{-\delta} (1-\tau)^{2\beta+1} F(\beta+1,\beta+1;2\beta+2;1-\tau), 
\label{eq:lum}
\end{equation}
where the constant $C$ is independent of $\tau$.   We will study the behavior of this luminosity first in the small-$\tau$ limit and then for $\tau$ of order unity.

For $\tau\ll 1$ we use the expansion
\bea
F(a,b;a+b;u) &=& \frac{\Gamma(a+b)}{\Gamma(a)\Gamma(b)} \sum\limits_{n=0}^{\infty} \frac{(a)_n (b)_n}{(n!)^2} 
\nn\\
&\ & \times\
\left[ 2 \psi(n+1) - \psi(a+n) - \psi(b+n) - \ln (1-u) \right] (1-u)^n\, ,
\label{eq:id3}
\eea
in terms of the Polchhammer symbol $(a)_n=\Gamma(a+n)/\Gamma(a)$.
For small $\tau$, we can approximate the hypergeometric function simply by the first term in its expansion Eq. \eqref{eq:id3}, and
\begin{equation}
{\cal L}(\tau) = C^2 \tau^{-\delta} \left( \ln \frac{1}{\tau} - 2\psi(\beta+1) - 2 \gamma_E \right)\, ,
\end{equation}
which implies, for the logarithmic derivatives of the luminosity, Eq. \eqref{eq:sn-def},
\begin{align}
s_1(\tau) &= \delta + \left( \ln \frac{1}{\tau} - 2\psi(\beta+1) - 2 \gamma_E \right)^{-1}, \nonumber \\
s_n (\tau) &= (n-1)! \left( \ln \frac{1}{\tau} - 2\psi(\beta+1) - 2 \gamma_E \right)^{-n}\, .
\end{align}
In the limit $\tau \to 0$, $s_1(\tau) \to \delta$ while $s_n(\tau)$ vanishes like $[\ln(1/\tau)]^{-n}$, confirming the validity of the single power approximation, even when the power $\delta$ is not large.

For large values of $\tau$, it is convenient to use the integral form of the luminosity function,
\begin{align}
{\cal L}(\tau) &\propto \int_\tau^1 \frac{dz}{z} z^{-\delta} (1-z)^\beta \left( \frac \tau z \right)^{-\delta} \left( 1-\frac \tau z \right)^\beta \nonumber \\
&= \tau^{-\delta} \int_\tau^1 \frac{dz}{z} (1-z)^\beta \left(1-\frac \tau z\right)^\beta\, ,
\end{align}
so that we have
\begin{align}
s_1 (\tau) &= -\frac{1}{{\cal L}(\tau)} \tau \frac{d}{d\tau} {\cal L}(\tau) \nonumber \\
&= \delta + \frac {\beta\, \tau \int_\tau^1 \frac{dz}{z^2} (1-z)^\beta \left(1-\frac \tau z\right)^{\beta-1}}{\int_\tau^1 \frac{dz}{z} (1-z)^\beta \left(1-\frac \tau z\right)^\beta}\, .
\label{eq:s_1-result}
\end{align}
In the large $\beta$ limit, both the integral in the numerator and the integral in the denominator can be performed using the saddle point approximation. Both integrands reach their maxima at approximately $z=\sqrt{\tau}$ and have nearly the same peak width. The ratio of the two integrals is then well approximated by the ratio of the integrands at $z=\sqrt{\tau}$,
\begin{align}
s_1 (\tau) & \approx \delta + \left. \frac{\beta\, \tau}{z\left(1-\frac \tau z\right)} \right|_{z=\sqrt{\tau}} \nonumber \\
&= \delta + \beta\,  \frac{\sqrt{\tau} }{(1-\sqrt{\tau})}\, ,
\end{align}
and
\begin{equation}
s_n (\tau) = \beta\, \left( \frac{d}{d\ln \tau} \right)^{n-1}  \frac{\sqrt{\tau} }{(1-\sqrt{\tau})}\, .
\end{equation}
Therefore, $s_1(\tau) = \delta + \beta\sqrt{\tau} + {\cal O}(\sqrt{\tau})$ and $(1/n!) \ s_n (\tau) / \left[ s(\tau) \right]^n = O (\beta^{-(n-1)}) << 1$, suggesting that the single power approximation is valid for  $\tau$ large as well as small.

\bibliographystyle{JHEP}
\bibliography{Tresum}

\end{document}